\documentclass[traditabstract]{aa}
\usepackage{txfonts}
\usepackage{graphicx}
\usepackage{float}
\usepackage{natbib}
\usepackage{hyperref}
\usepackage[english]{babel}
\usepackage{color}
\usepackage{amsmath}

\title{The redshift evolution of the distribution of star formation among dark matter halos as seen in the infrared}

\author{Matthieu B{\'e}thermin\inst{1} \and Lingyu Wang\inst{2} \and Olivier Dor\'e\inst{3,4} \and Guilaine Lagache\inst{5} \and Mark Sargent\inst{1} \and Emanuele Daddi\inst{1} \and  Morgane Cousin\inst{5} \and Herv\'e Aussel\inst{1}}
\institute{Laboratoire AIM-Paris-Saclay, CEA/DSM/Irfu - CNRS - Universit\'e Paris Diderot, CEA-Saclay, pt courrier 131, F-91191 Gif-sur-Yvette, France, email: matthieu.bethermin@cea.fr \and Computational Cosmology, Department of Physics, University of Durham, South Road, Durham, DH1 3LE, UK \and Jet Propulsion Laboratory, California Institute of Technology, 4800 Oak Grove Drive, Pasadena, California, U.S.A \and California Institute of Technology, MC249-17, Pasadena, CA 91125 USA \and Institut d'Astrophysique Spatiale, CNRS (UMR8617) Universit\'e Paris-Sud 11, B\^atiment 121, Orsay, France}

\date{Received 12 April 2013 / Accepted 8 June 2013}

\abstract{Recent studies have revealed a strong correlation between the star formation rate (SFR) and stellar mass of the majority of star-forming galaxies, the so-called star-forming main sequence. An empirical modeling approach (the 2-SFM framework) that distinguishes between the main sequence and rarer starburst galaxies is capable of reproducing most statistical properties of infrared galaxies, such as number counts, luminosity functions, and redshift distributions. In this paper, we extend this approach by establishing a connection between stellar mass and halo mass with the technique of abundance matching. Based on a few simple assumptions and a physically motivated formalism, our model successfully predicts the (cross-)power spectra of the cosmic infrared background (CIB), the cross-correlation between CIB and cosmic microwave background (CMB) lensing, and the correlation functions of bright, resolved infrared galaxies measured by \textit{Herschel}, \textit{Planck}, ACT, and SPT. We use this model to infer the redshift distribution of CIB-anisotropies and of the CIBxCMB lensing signal, as well as the level of correlation between CIB-anisotropies at different wavelengths. We study the link between dark matter halos and star-forming galaxies in the framework of our model. We predict that more than 90\% of cosmic star formation activity occurs in halos with masses between $10^{11.5}$ and $10^{13.5}$\,M$_\odot$. If taking subsequent mass growth of halos into account, this implies that the majority of stars were initially (at $z>$3) formed in the progenitors of  clusters ($M_h(z=0)>10^{13.5}\,M_\odot$), then in groups ($10^{12.5}<M_h(z=0)<10^{13.5}\,M_\odot$) at $0.5<z<3$, and finally in Milky-Way-like halos ($10^{11.5}<M_h(z=0)<10^{12.5}\,M_\odot$) at $z<0.5$. At all redshifts, the dominant contribution to the star formation rate density stems from halos of mass $\sim10^{12}\,$M$_\odot$, in which the instantaneous star formation efficiency --  defined here as the ratio between SFR and baryonic accretion rate -- is maximal ($\sim$70\%). The strong redshift-evolution of SFR in the galaxies that dominate the CIB is thus plausibly driven by increased accretion from the cosmic web onto halos of this characteristic mass scale. Material (effective spectral energy distributions, differential emissivities of halos, relations between $M_h$ and SFR) associated to this model is available at \url{http://irfu.cea.fr/Sap/Phocea/Page/index.php?id=537}.}

\keywords{Galaxies: star formation -- Galaxies: halos -- Galaxies: statistics -- Cosmology: diffuse radiation -- Cosmology: dark matter -- Submillimeter: galaxies}

\titlerunning{The redshift evolution of the distribution of star formation among dark matter halos as seen in the infrared}

\authorrunning{B\'ethermin et al.}

\begin{document}

\maketitle

\section{Introduction}

A detailed understanding of galaxy formation in the cosmological context is one of the main problems of modern astrophysics. The star formation rate (SFR) of galaxies across cosmic time is one of the key observables to understand their evolution. However, measurements of SFR are difficult, because most of the UV light emitted by young massive stars is absorbed by interstellar dust. This light is reradiated in the infrared between 6\,$\mu$m and 1\,mm. The cosmic infrared background (CIB), detected for the first time in FIRAS data \citep{Puget1996,Fixsen1998,Hauser1998}, is the relic of all dust emissions since the recombination. It is the strongest background after the cosmic microwave background (CMB), and it contains half of the energy emitted after recombination \citep{Hauser2001,Dole2006}. Identifying the sources responsible for this background and their physical properties is thus crucial to understanding the star formation history in the Universe. Unfortunately, because of the confusion caused by the limited resolution of the current infrared/millimeter facilities \citep{Condon1974,Dole2004}, only a small fraction of this background can be directly resolved into individual sources at wavelengths longer than 250\,$\mu$m \citep{Oliver2010,Bethermin2012b}, where the CIB becomes dominated by z$>$1 sources \citep{Lagache2003,Lagache2005,Bethermin2011,Bethermin2012b}. We thus have to study the statistical properties of the unresolved background, if we want to unveil the infrared properties of galaxies that host the bulk of the obscured star formation at high redshift.\\

We can use statistical tools to measure the photometric properties of galaxies emitting the CIB. P(D) analysis \citep{Condon1974,Patanchon2009} is a method measuring the flux distribution of sources below the confusion limit by considering only the pixel histogram of an infrared/millimeter map. \citet{Glenn2010} has managed to measure the number counts (flux distribution) of SPIRE sources down to one order of magnitude below the confusion limit using P(D), and they resolve about two-thirds of the CIB into individual sources. Strong constraints on contributions to the CIB were also derived by stacking analyses \citep{Dole2006,Marsden2009}. This method allows measuring the mean flux of a population individually detected at a shorter wavelength\footnote{24\,$\mu$m is often used because $\sim$ 80\% of the background is resolved into sources at this wavelength \citep{Papovich2004,Bethermin2010a}.}, but not in the far-infrared/millimeter, by stacking cutout images centered on short-wavelength detections. Number counts below the confusion limit were measured with a method based on stacking \citep{Bethermin2010a,Bethermin2010b}. In addition to this, \citet{Bethermin2012b} also measured counts per redshift slice using an input catalog containing both 24\,$\mu$m fluxes and redshifts and a complex reconstruction of the counts based on stacking. These analyses provided constraints on the CIB redshift distribution. Nevertheless, an empirical model is still needed to deduce the obscured star formation history from number counts at various wavelengths \citep[e.g.][]{Le_Borgne2009,Valiante2009,Franceschini2009,Bethermin2011,Marsden2010,Rahmati2011,Lapi2011,Gruppioni2011}.\\
 
Large-scale CIB anisotropies measured by \textit{Spitzer} \citep{Lagache2007,Grossan2007,Penin2012b}, BLAST \citep{Viero2009}, the South Pole Telescope \citep[SPT,][]{Hall2010}, \textit{Herschel} \citep{Amblard2011,Viero2012}, and \textit{Planck} \citep{Planck_CIB} also provide degenerate constraints on the evolution of infrared-galaxy emissivities and the link between infrared galaxies and dark matter halos. This degeneracy can be broken by combining anisotropy information with infrared number counts (see above). The first generation of models used to predict/interpret CIB anisotropies was based on a combination of an evolutionary model of emissivities of infrared galaxies and a linear bias or a halo occupation distribution (HOD) model describing the spatial distribution of galaxies \citep{Knox2001,Lagache2007,Amblard2007,Viero2009,Hall2010,Planck_CIB,Amblard2011,Penin2012a,Xia2012}. The emissivities are deduced from a model of galaxy evolution \citep[e.g.][]{Bethermin2011,Lapi2011} or represented by a simple parametric function \citep{Hall2010,Amblard2011}. However, these models assume that there is no dependency between clustering and luminosity and in general a single HOD or linear bias for all redshifts. Consequently, these models have difficulty fitting all wavelengths simultaneously. \citet{Shang2012} propose a new approach assuming an infrared-light-to-mass ratio that varies with halo mass and redshift (see also the \citealt{DeBernardis2012} approach, which focused on 250\,$\mu$m). This new model is also able to roughly reproduce the number counts (LFs respectively), though their description of infrared galaxies is simplistic (a single SED for all galaxies at all redshifts, no scatter on the mass-to-light ratio). Another approach was proposed by \citet{Addison2013}, who combine a backward-evolution counts model that is very similar to the \citet{Bethermin2011} approach and a scale-dependent effective bias of infrared galaxies to predict the CIB power spectrum. This simplified approach is very efficient in fitting the data, but is purely descriptive and provides little information on the physical link between galaxies and dark matter halos.\\

In this paper, we propose a new approach to modeling both CIB anisotropies and galaxy number counts based on the observed relation between physical properties in galaxies and their evolution with redshift. We use the stellar mass (M$_\star$) as a gateway to link the halo mass (M$_h$) and star formation rate (SFR). The stochastic link between SFR and M$_\star$ is modeled following the \citet{Bethermin2012c} model (hereafter B12), which is based on the observed main sequence (MS) of star-forming galaxies (i.e., a strong correlation between stellar mass  and star formation rate evolving with redshift, \citealt{Noeske2007,Elbaz2007,Daddi2007,Rodighiero2011}) and provides one of the best fits of mid-infrared-to-radio number counts. It contains two types of star-forming galaxies with different spectral energy distributions (SEDs): secularly star-forming MS galaxies and a population of episodic, probably merger-driven starbursts with a strong excess of SFR compared to the main sequence following the two-star-formation-mode (2SFM) formalism introduced in \citet{Sargent2012}. The relation between stellar and halo mass is derived by abundance matching \citep{Vale2004, Conroy2009,Behroozi2010} and assuming a monotonic relation without scatter between these quantities. \citet{Bethermin2012a} used this technique to connect SFR and M$_h$ for MS galaxies in a qualitative way (see also \citealt{Wang2012}). This paper improves and extends the approach of \citet{Bethermin2012a} in order to predict the anisotropy of the CIB and the clustering of infrared galaxies. This new formalism also permits us to describe the quenching of star formation in massive galaxies and their satellites in a phenomenological way and involves a refined treatment of subhalos.\\

In Sect.\,\ref{sect:philo}, we present the philosophy of our approach. In Sect.\,\ref{sect:idea}, we explain the ideas on which our model is based, especially how we assign infrared properties to galaxies hosted by a dark matter halo. In Sects.\,\ref{sect:formalism_cib} and \ref{sect:formalism_acf}, we describe the formalism used to compute the power spectrum of CIB anisotropies and angular correlation functions of infrared galaxies, respectively. In Sect.\,\ref{sect:results}, we present the results of our modeling and an extensive comparison with observations. In Sect.\ref{sect:validity}, we discuss the successes, but also the limitations, of our model. In Sect.\,\ref{sect:origin}, we describe the properties of the CIB predicted a posteriori by our model, such as redshift distribution or correlation between bands. In Sect.\,\ref{sect:sfh}, we discuss how the history of star formation history proceeds depending on the mass of dark matter halos. We conclude in Sect.\,\ref{sect:conclusion}. In Appendix\,\ref{sect:conv}, we provide tables of conversion from multipole $\ell$ to angle $\theta$ and from frequencies to wavelengths, since we use both conventions interchangeably in the paper.\\

In this paper, we assume a WMAP-7 cosmology \citep{Larson2010} and a \citet{Chabrier2003} initial mass function.\\

\section{Philosophy of our approach}

\label{sect:philo}

The majority of previous CIB models were purely phenomenological, and they describe the emissivity and clustering of infrared galaxies using a large set of free parameters. This approach is useful for deriving quantities, such the mass where star formation is most efficient. However, it is sometimes hard to test the validity of these models since a good fit can be obtained easily considering their number of free parameters. \citet{Kim2012} proposes a physical approach based on a semi-analytical galaxy formation model, which unfortunately has substantial discrepancies with the data. We propose an alternative phenomenological approach, which represents an intermediate solution between a fully empirical and a physical model. We minimize the number of free parameters and build our analysis on the observed relation between physical quantities (e.g., the specific star formation rate in main-sequence galaxies calibrated from optical, near-infrared, far-infrared, and radio data). We thus do not aim to fine tune the various parameters of the model, but rather to test whether the scaling laws measured from external datasets (measurements on small samples extrapolated to all galaxies, optical/near-infrared measurements of stellar mass, or SFR) are compatible with the data under different scenarios. For this reason, we chose an approach based on abundance matching with no free parameters to describe the link between stellar mass and halo mass and use the B12 model, which follows the same philosophy to link infrared properties and stellar mass.\\

\begin{figure*}
\centering
\includegraphics{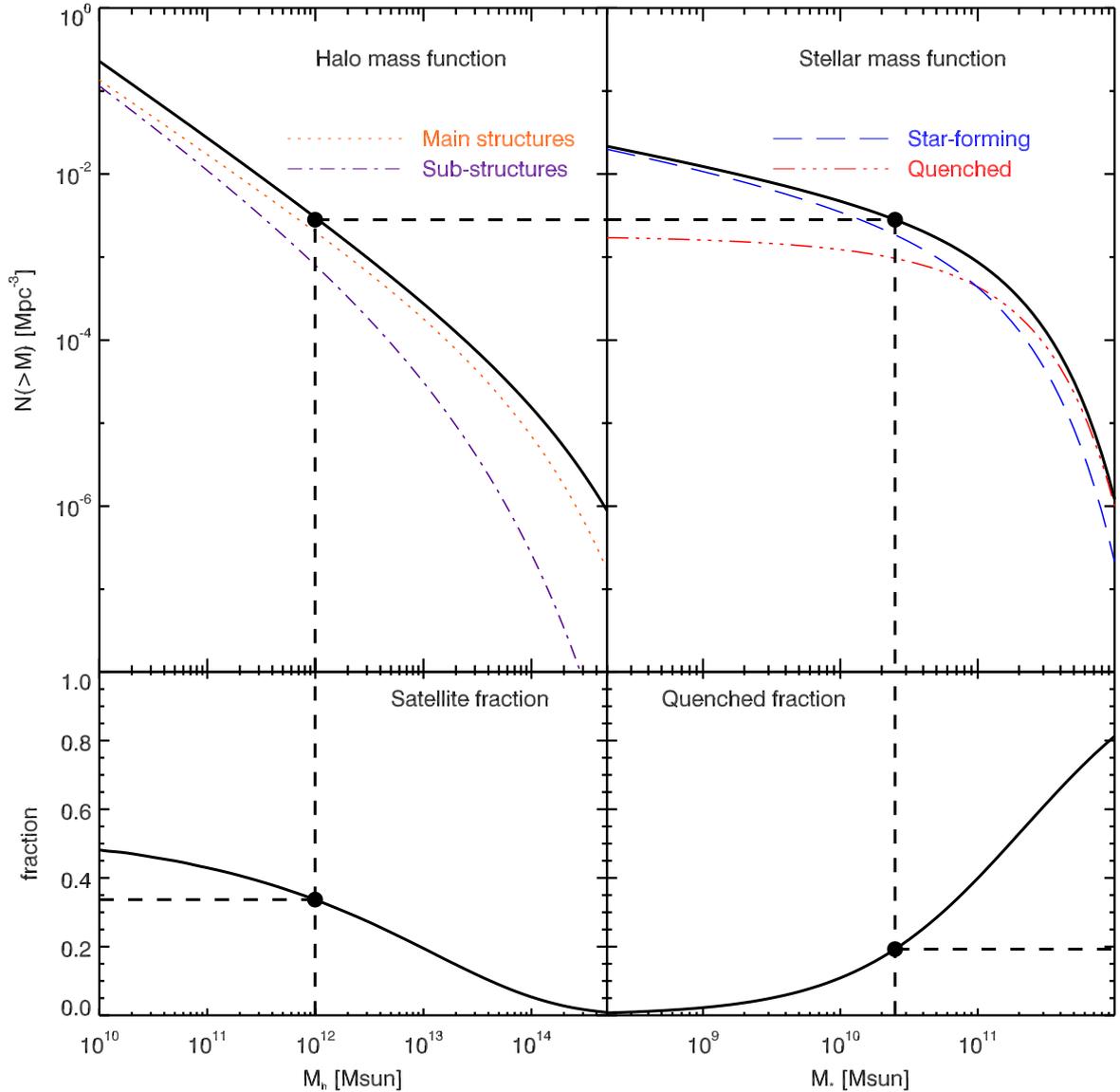}
\caption{\label{fig:am} Illustration of the method used to connect various quantities by abundance matching as described in Sect.\,\ref{sect:am}. The dashed line illustrates the connection between various quantities. We arbitrarily chose to plot  results at $z=0.5$. \textit{Upper left panel}: integral HMF and contribution of main halos (\textit{dotted orange line}) and subhalos (\textit{dot-dashed purple line}). \textit{Upper right panel}: integral SMF and contribution of star-forming (\textit{long-dashed blue line}) and quenched (\textit{three-dot-dashed red line}). \textit{Lower left panel}: variation with halo mass of the fraction of halos that are subhalos. \textit{Lower right panel}: fraction of quenched galaxies as a function of stellar mass.}
\end{figure*}

\section{Connecting star formation and halo mass by abundance matching}

\label{sect:idea}

In this section, we describe how we stochastically assign properties of star formation to galaxies as a function of the host halo mass by combining prescriptions from previous models. In Sects.\,\ref{sect:hmf} and \ref{sect:smf}, we present the halo and stellar mass functions used in this paper. We then describe how we connect stellar and halo mass by abundance matching in Sect.\,\ref{sect:am}. Section\,\ref{sect:sfrmass} describes how we stochastically attribute a star-formation rate from the stellar mass using recipes based on the B12 model. Finally, we describe how we deduce infrared properties of the galaxies from their physical properties in Sect.\,\ref{sect:irprop} using the B12 model.

\subsection{Halo mass function}

\label{sect:hmf}

In our analysis, we used the halo mass function (HMF) of \citet{Tinker2008} (in our notation $\frac{d^2 N}{d\textrm{log}(M_h) dV}$). We chose halo mass (M$_h$) to be defined by an overdensity of $\Delta = 200$ (often called M$_{200}$). This HMF evolves with redshift and was calibrated on N-body simulations. We also need the mass function of subhalos, which are supposed to host satellite galaxies. The one we adopt here comes from \citet{Tinker2010}, and provides the mass distribution of subhalos in a parent halo of total mass $M_h$:
\begin{equation}
\begin{split}
\frac{dN}{d\textrm{log}(m_{\rm sub}) \times \textrm{ln}(10)}(m_{\rm sub}|M_h) = \\
0.30 \times \left ( \frac{m_{\rm sub}}{M_{\rm h}} \right)^{-0.7} \times  \textrm{exp} \left ( -9.9 \left ( \frac{m_{\rm sub}}{M_{\rm h}} \right )^{2.5} \right),
\end{split}
\end{equation}
where $m_{\rm sub}$ is the subhalo mass. In our analysis we neglect sub-structures inside subhalos. The mass function of subhalos is thus
\begin{equation}
\begin{split}
\frac{d^2 N}{d\textrm{log}({m_{\rm sub}}) dV}(m_{\rm sub}) = \\
\int_0^{M_h} \frac{dN}{d\textrm{log}(m_{\rm sub})}(m_{\rm sub} = \mathcal M_h|M_h) \times \frac{d^2N}{d\textrm{log}(M_h) dV} d\textrm{log}{M_h},
\end{split}
\end{equation}
We also introduce a pseudo "total" mass function, given by the sum of the mass functions of halos and subhalos:
\begin{equation}
\begin{split}
\frac{d^2 N}{d\textrm{log}({\mathcal M_h}) dV}({\mathcal M_h}) = \frac{d^2 N}{d\textrm{log}(M_h) dV}(M_h = {\mathcal M_h})\\
+ \frac{d^2 N}{d\textrm{log}({m_{\rm sub}}) dV}(m_{\rm sub} = {\mathcal M_h}).
\end{split}
\end{equation}
Here ${\mathcal M_h}$ may stand for either the total mass of a halo ($M_h$) for a main structure or the mass of the subhalo for a substructure ($m_{\rm sub}$). This function will be useful for our abundance-matching procedure presented Sect\,\ref{sect:am}, because we assume that the properties of galaxies are linked with ${\mathcal M_h}$. However, this is not exactly a mass function, because subhalos are counted twice, namely both in the total mass function and in the subhalo mass function. Figure\,\ref{fig:am} (lefthand column) shows the contribution of main and subhalos at z$\sim$0.5. The majority of high-mass halos are main halos, while a large fraction of low-mass halos are substructures of more massive halos. At $\mathcal M_h$=10$^{12}$\,M$_\odot$, $\sim$1/3 of halos are subhalos of more massive halos.\\

\subsection{Stellar mass function}

\label{sect:smf}

We used the same stellar mass function (SMF) of star-forming galaxies as B12, in order to be consistent with this model, which is used to link the stellar mass to the infrared properties (see Sect.\,\ref{sect:sfr_Mh}). This mass function is parametrized by an evolving \citet{Schechter1976} function:
 \begin{equation}
\phi = \frac{dN}{d \textrm{log}(M_\star) dV} = \phi_b (z) \times \left ( \frac{M_\star}{M_b} \right )^{- \alpha_{MF}} \times \textrm{exp} \left ( -  \frac{M_\star}{M_b} \right ) \times  \frac{M_\star}{M_b} \textrm{ln}(10),
\end{equation}
with a redshift-invariant characteristic mass $M_b$ and faint-end slope $\alpha_{MF}$. The characteristic density $\phi_b$ is constant between z=0 and z=1 but decreases at z$>$1 as
\begin{equation}
\textrm{log}(\phi_b) = \textrm{log}(\phi_b)(z<1) + \gamma_{SFMF} (1-z).
\end{equation}
The various parameters were chosen to reproduce the observed evolution of the mass function of star-forming galaxies. Their values are given in B12. We checked that our results are not modified significantly if we use the double-Schechter fits of the measured SMF of \citet{Ilbert2013} instead of this simplified parametric form.\\

To correctly populate dark matter halos, we also need to account for the population of non-star-forming galaxies (called quenched galaxies hereafter), which are essentially red, passively-evolving, elliptical galaxies below the main sequence. Star formation activity in these objects is weak, and this population was thus ignored in the B12 model because of their negligible infrared emission. These galaxies do, however, contribute significantly to the mass function at high mass (e.g. \citealt{Ilbert2013}, see Fig.\,\ref{fig:am} upper righthand panel) and thus are generally the kind of galaxy that is encountered in massive halos. We used the mass function of quiescent galaxies from \citet{Ilbert2010} at z$<$2 and \citet{Ilbert2013} at z$>$2, fitted by a \citet{Schechter1976} function. log($\phi_b$), log(M$_b$), and $\alpha$ are interpolated between the center of each redshift bins and extrapolated at $z>3$. The total mass function (showed Fig.\,\ref{fig:am} upper right panel) is the sum of the contribution of both quenched and star-forming galaxies. The fraction of quenched galaxies at a given stellar mass and redshift is called $q(M_\star,z)$. The upper righthand panel of Fig.\,\ref{fig:am} shows the mass function and its decomposition into quenched and star-forming galaxies. The lower right panel shows the fraction of quenched galaxies as a function of stellar mass. At high mass ($M_\star>10^{11}$\,M$_\odot$) and low redshift (z$<1$), the majority of galaxies are quenched, when the other regimes are dominated by star-forming galaxies.\\

\begin{figure}
\centering
\includegraphics{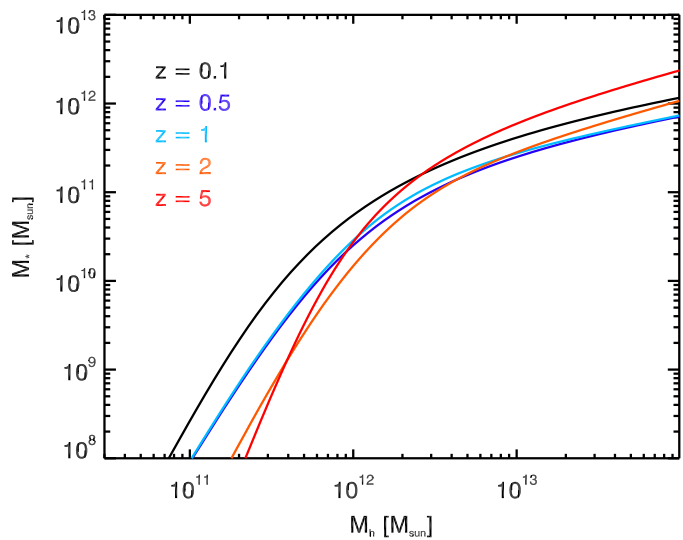}
\caption{\label{fig:mstarmh} Relation between the halo mass and the stellar mass at various redshifts found by our abundance matching procedure described in Sect.\,\ref{sect:am}.}
\end{figure}

\subsection{Connecting stellar mass and halo mass through abundance matching}

\label{sect:am}

The abundance-matching technique is based on the hypothesis of a monotonic link between two quantites. This is a fair assumption for the link between the stellar and halo mass of a central galaxy at any given redshift \citep{More2009,Moster2010,Behroozi2010}. In this work, we follow \citet{Behroozi2010} and \citet{Watson2013} by assuming that subhalos and main halos follow the same $M_\star$-$M_h$ relation. Under this assumption, we associate halo mass to a stellar mass by putting the n-th most massive galaxy (in term of stellar mass) into the n-th most massive halo. In practice, we do not use catalogs, but rather analytic mass functions. The nonparametric function linking stellar and halo mass ($M_\star=f(\mathcal M_h)$) is thus the solution of the implicit equation
\begin{equation}
n_{M_\star}(>f(\mathcal M_h)=M_\star) = n_{\mathcal M_h},
\end{equation}
where $n_{M_\star}(>M_\star)$ is the number density (in comoving units) of galaxies more massive than $M_\star$ (i.e., the integral of the mass function), and $n_{\mathcal M_h}$ the equivalent for halo mass. We can thus associate halo mass to stellar mass by taking the halo mass at which the number density of galaxies and halos are the same, as illustrated by Fig.\,\ref{fig:am} (upper panels). Figure\,\ref{fig:mstarmh} shows the resulting M$_\star$-$\mathcal M_h$ relation, which evolves little with redshift and displays a break at $\mathcal M_h \sim 10^{12}$\,M$_\odot$, as classically found in the literature \citep[e.g.,][]{Moster2010,Behroozi2010,Leauthaud2012}.\\

In this approach, we neglect the effect of the scatter on the stellar-to-halo-mass ratio, which would further complicate our analysis of CIB anisotropies. This induces a bias on the estimation of $f$. However, this effect is smaller than the statistical uncertainties for $M_h>10^{14.5}$\,M$_\odot$ \citep{Behroozi2010}, which host mainly passive galaxies (see Fig.\,\ref{fig:am}). The scatter around $f$ could also induce a bias on the estimate of the observables as the CIB anisotropies. However, the large-scale anisotropies are sensitive to the mean emissivity of galaxies and are not affected by the scatter. The small-scale Poisson term can be computed directly from a count model without assumptions on the dark matter (see Sect.\,\ref{sect:formalism_cib}). The angular correlation of bright resolved galaxies can be affected by the scatter on halo mass\label{sect:formalism_acf}, but the scatter between stellar and halo mass ($\sim0.15$\,dex) has the same order of magnitude as the scatter between stellar mass and star formation rate, which is taken into account by our model ($\sim$0.15-0.2\,dex). The impact of the scatter of the stellar mass-halo mass relation on the correlation function is thus expected to be relatively small.\\   

\subsection{Connecting star formation rate to stellar mass}

\label{sect:sfrmass}

\label{sect:sfr_Mh}

In the previous section, we explained how we can assign a stellar mass to a galaxy knowing its halo mass. Unfortunately, we cannot link star formation rate to stellar mass by abundance matching by assuming a monotonic relation. This hypothesis is only valid for main-sequence galaxies, but not for quenched ones, for which sSFR$<<$sSFR$_{\rm MS}$, and starburst galaxies, for which sSFR$>>$sSFR$_{\rm MS}$, where sSFR is the specific star formation rate; i.e., SFR/M$_\star$, and sSFR$_{\rm MS}$ is the typical value of this parameter in main-sequence galaxies. For quenched galaxies, we neglect the star formation and thus take SFR=0\,M$_\odot$.yr$^{-1}$ for simplicity. For star-forming galaxies (main-sequence and starburst), we assume that SFR follows a double log-normal distribution at fixed redshift and stellar mass \citep[][B12]{Sargent2012}:
\begin{equation}
\label{eq:ssfr_distrib}
\begin{split}
p\left (\textrm{log(SFR)}|M_\star,z \right ) \propto p_{\rm MS} + p_{\rm SB} \\
\propto \textrm{exp} \left ( -\frac{ \left (\textrm{log(SFR)} -  \textrm{log(sSFR}_{\rm MS}\times M_\star) \right )^2}{2 \sigma_{\rm MS}^2} \right ) \\
+ r_{SB} \times \textrm{exp} \left (-\frac{ \left ( \textrm{log(SFR)}- \textrm{log(sSFR}_{\rm MS}\times M_\star) - B_{\rm SB}) \right )^2}{2 \sigma_{\rm SB}^2} \right ).\\
\end{split}
\end{equation}
The first term describes the sSFR distribution of main-sequence galaxies ($p_{\rm MS}$), and the second one that of starbursts ($p_{\rm SB}$). Star formation in starburst is boosted on average by a factor $ B_{SB}$\footnote{A detailed discussion of the SFR-enhancements of starbursts and their description in the 2-SFM framework is provided in \citet{Sargent2013}.}, and $\sigma_{\rm MS}$ and $\sigma_{\rm SB}$ are the dispersions around the central values. The evolution of the main-sequence is parametrized as in B12:  
\begin{equation}
\begin{split}
\textrm{sSFR}_{\rm MS} (z,M_\star) = & \textrm{sSFR}_{\rm MS,0} \times \left ( \frac{M_\star}{10^{11}\,M_\odot} \right)^{\beta_{\rm MS}} \\
	& \times  \left ( 1+ \textrm{min} \left (z, z_{\rm evo} \right ) \right )^{\gamma_{MS}}.
\end{split}
\end{equation}
The starburst ratio $r_{\rm SB}$ (i.e., the relative normalization of the two log-normal distributions) is also provided by the B12 model:
\begin{equation}
r_{SB}(z) = r_{SB,0} \times \left ( 1+ \textrm{min} \left (z, z_{SB} \right ) \right )^{\gamma_{SB}}, \, \textrm{where} \, z_{SB} = 1.
\end{equation}
\\

\begin{figure}
\centering
\includegraphics[width=8.5cm]{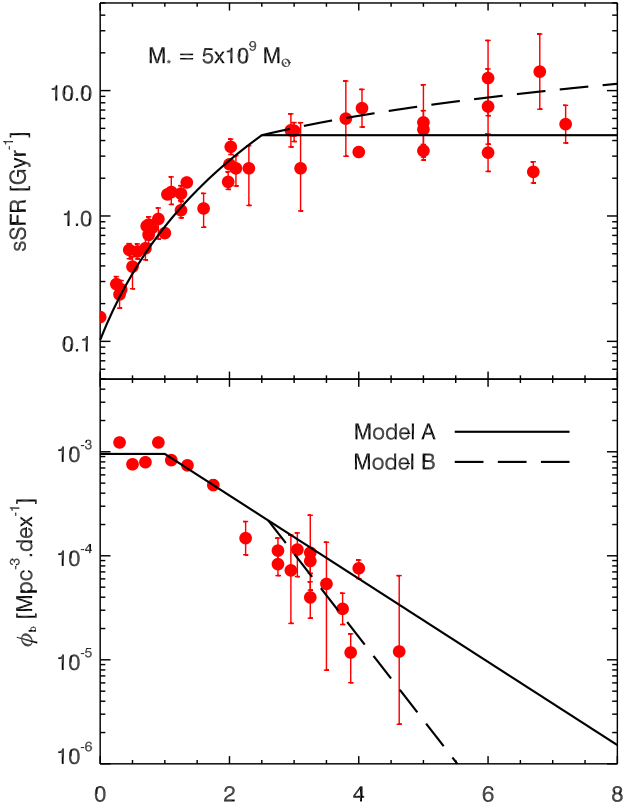}
\caption{\label{fig:evo} Upper panel: Evolution of specific star formation rate in main-sequence galaxies of $5\times10^9\,M_\odot$. \textit{Lower panel:} Evolution of the characteristic density $\phi_b$ of the mass function with redshift. \textit{Both panel:} model A (B12 model) is represented with a solid line and model B (modified version with higher sSFR and lower density at high redshift) with a long-dashed line. We use the compilation of data points of \citet{Sargent2012}.}
\end{figure}

The values of the parameters in our base model (model A) are the ones provided in B12\footnote{Parameters provided in B12 are given assuming a \citet{Salpeter1955} IMF when this paper assumes a \citet{Chabrier2003} IMF. A correction of 0.24\,dex thus has to be applied to some of the parameters to take this difference in IMF into account.}. We also used a second version of the model (model B) for which the high redshift trend was slightly modified, following the findings of a slowly increasing sSFR at high redshift \citep{Stark2012,deBarros2012,Gonzalez2012}. These higher values are the consequence of an improved  modeling of the contribution of nebular line emissions to flux in near-infrared broad-band filters. For this modified version, we assume an evolution of sSFR at $z>2.5$ in $(1+z)$. To avoid overpredicting bright millimeter counts, we compensate this increase of sSFR by a quicker decrease in the characteristic density of the stellar mass function in the same redshift range ($(1+z)^{-0.8}$ instead of $(1+z)^{-0.4}$), thus keeping the same number of bright objects. As shown in Fig.\,\ref{fig:evo}, both these scenarios are compatible with the data, because of the large scatter on the measurements. In Sects.\ref{sect:results} and \ref{sect:validity}, we discuss which scenario is actually favored by infrared observations.\\

\subsection{Infrared outputs of galaxies}

\label{sect:irprop}

In massive galaxies, the bulk of the UV light coming from young stars is absorbed by dust and re-emitted in the infrared. We can thus assume that the bolometric infrared  (8-1000\,$\mu$m) luminosity $L_{\rm IR}$ is proportional to star formation rate (\citealt{Kennicutt1998}, the conversion factor is $K = SFR/L_{\rm IR} = 1\times10^{-10}$\,M$_\odot$yr$^{-1}$L$_\odot^{-1}$ if we assume a \citet{Chabrier2003} IMF). In low-mass galaxies, a significant part of the UV light escapes from the galaxy and infrared emission is no longer proportional to SFR. Star formation rates can then  be estimated from an uncorrected UV and an infrared component ($SFR = SFR_{UV}+SFR_{IR}$). The infrared luminosity is then given by
\begin{equation}
\label{eq:att}
L_{IR} = \frac{\textrm{SFR}_{IR}}{K} = \frac{\textrm{SFR}}{K} \times \frac{10^{0.4 \times r_{\rm UV}}}{1+10^{0.4 \times r_{\rm UV}}} = \frac{\textrm{SFR}}{K} \times g(M_\star),
\end{equation}
where $r_{\rm UV}$ is the ratio between obscured and unobscured star formation. We use the r$_{\rm UV}$-M$_\star$ relation of \citet{Pannella2009} to compute $g(M_\star)$:
\begin{equation}
r_{\rm UV} = 2.5 \textrm{log} \left ( \frac{\textrm{SFR}_{IR}}{\textrm{SFR}_{UV}} \right ) = 4.07 \times \textrm{log}\left(\frac{M_\star}{M_\odot} \right) - 39.32.
\end{equation}
$g(M_\star)$ tends to 0 at low mass and 1 at high mass. The UV light from young stars thus totally escapes the low-mass galaxies, but is fully reprocessed and emitted as infrared radiation in massive galaxies. This is due to a larger amount of dust in massive galaxies, which causes a greater attenuation.\\

We used the same spectral energy distribution (SED) templates as B12, based on \citet{Magdis2012}. There are different templates for main-sequence and starburst galaxies, both of which are assumed to evolve with redshift. We did not adopt a single template for a given type of galaxy at a given redshift, but assumed a scatter on the mean interstellar radiation field $\langle U \rangle$, following B12. These SEDs were calibrated using $z<2$ data, which show a rise in the (rest-frame) dust temperature with redshift. At higher redshift, we assume no evolution. This assumption is discussed in Sect.\,\ref{sect:validity}. In this paper, we neglect the contribution of active galactic nuclei, which is only significant in the mid-infrared (B12).\\

\begin{figure}
\centering
\includegraphics{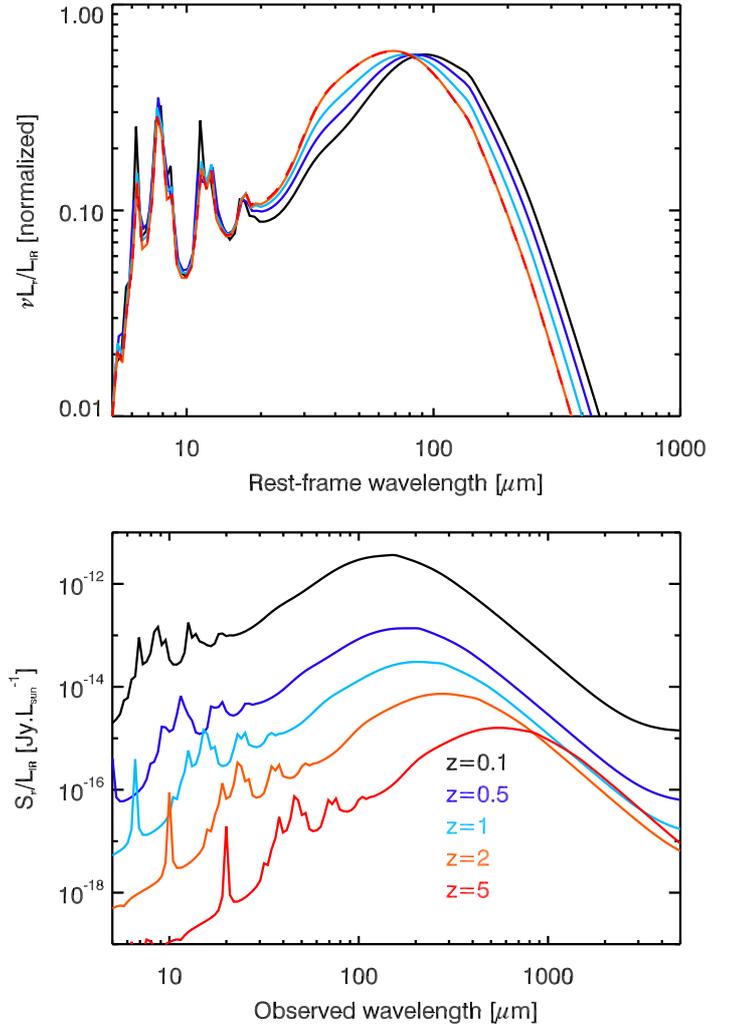}
\caption{\label{fig:sedeff} Effective SEDs of infrared galaxies used in our model at various redshifts. The upper panel shows the SEDs in $\nu L_\nu$ units normalized at $L_{\rm IR}=1\,L_\odot$ as a function of rest-frame wavelength. The lower panel shows the ratio between the observed flux density and $L_{\rm IR}$ as a function of observed wavelength.}
\end{figure}

\begin{figure}
\centering
\includegraphics{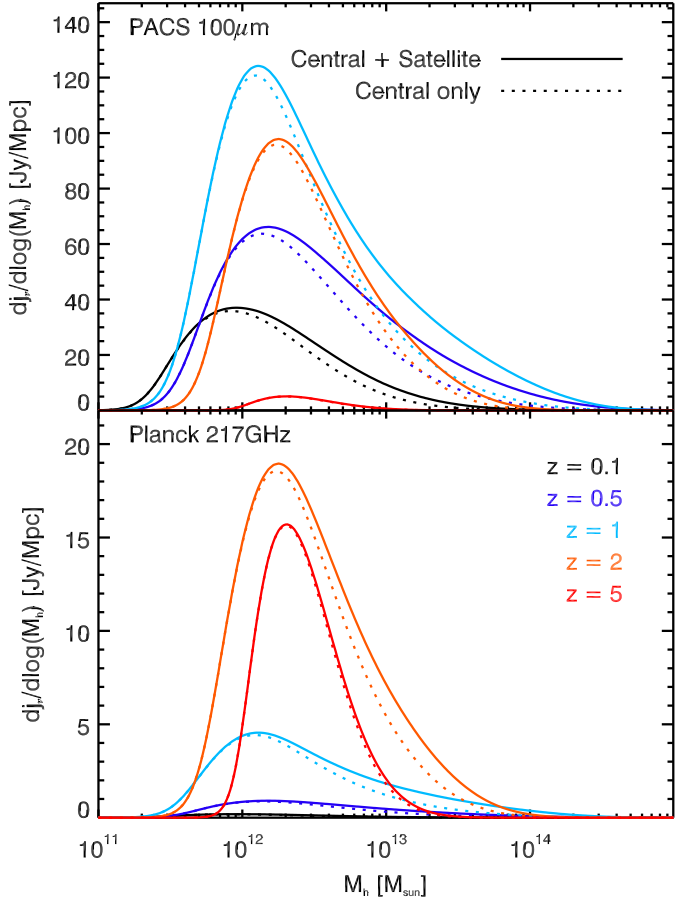}
\caption{\label{fig:jnu} Differential emissivities at 100\,$\mu$m (3000\,GHz) and 1.38\,mm (217\,GHz) as a function of halo mass at various redshifts predicted by model C. The solid lines are the contributions of all the galaxies to the infrared flux of a main halo, when the dotted lines indicate only the contribution of central galaxies.}
\end{figure}

\section{Computing CIB power spectrum}

\label{sect:formalism_cib}

We aim to compare the CIB anisotropies predicted by our model to observations in order to test its validity. This section presents the formalism used to derive the power spectrum (cross-spectrum) of the CIB at a given waveband (between two wavebands, respectively). One of the key benefits of the relation we established between SFR and ${\mathcal M_h}$ is that we can then rely on the well-known clustering properties of dark matter halos to predict the clustering of star-forming galaxies, and thus of CIB anisotropies. We use a method similar to \citet{Shang2012}. However, we modified their formalism to obtain a more natural notation and avoid renormalization of all terms by the total emissivity of infrared galaxies at a given redshift.\\

\subsection{Mean infrared emissivities of dark matter halos}

One of the key ingredients to compute is the mean emissivities of the halos. Classical CIB models assume that clustering and emissivity are independent, so they compute the total emissivity of galaxies at a given redshift. This approximation is not exact, and both emissivity and clustering vary with halo mass (see e.g. \citealt{Bethermin2012a}). We thus introduce the differential emissivity $dj_\nu/d\textrm{log}(M_h)$ of dark matter halos as a function of halo mass. This differential emissivity is the sum of the contribution of central galaxy and satellite galaxies:
\begin{equation}
\frac{dj_\nu}{d\textrm{log}(M_h)}(M_h,z) = \frac{dj_{\nu,\textrm{c}}}{d\textrm{log}(M_h)}(M_h,z)+ \frac{dj_{\nu,\textrm{sub}}}{d\textrm{log}(M_h)}(M_h,z).
\end{equation}
\\

The contribution of central galaxies to the differential emissivity is computed from mean infrared flux of galaxies hosted by a halo of mass $M_h$ and the HMF:
\begin{equation}
\begin{split}
\frac{dj_{\nu,\textrm{c}}}{d\textrm{log}(M_h)}(M_h,z)  = \frac{d^2N}{d\textrm{log}(M_h) dV} \times D_c^2 (1+z)\\
\times \frac{\textrm{SFR}_{\rm MS}(M_\star=f(M_h),z)}{K} \times g(M_\star=f(M_h),z)\\
\times s_\nu^{\rm eff}(z) \times \left (1-q \left ( M_\star=f \left ( M_h \right ),z \right ) \right ),
\end{split}
\end{equation}
where $D_c$ is the comoving distance and $s_\nu^{\rm eff}(z)$ is the effective SED of infrared galaxies at a given redshift; i.e., the mean flux density received from a population of star-forming galaxies with a mass corresponding to a mean infrared luminosity of 1\,L$_\odot$ (see Fig.\,\ref{fig:sedeff}), and $q \left ( M_\star=f \left ( M_h \right ),z \right )$ is the fraction of quenched galaxies (i.e. non-star-forming galaxies) as defined in Sect.\,\ref{sect:smf}. Because the shape of sSFR distribution is invariant with mass in our model, the mean flux density coming from a more massive population can thus be obtained just by rescaling this SED and taking the attenuation ($g(M_\star)$ defined in Eq.\,\ref{eq:att}) and the SFR-M$_\star$ relation into account. The effective SED is thus the mean of each type of SED weighted by their contributions to the background (provided by Eq.\,\ref{eq:ssfr_distrib}):
\begin{equation}
s_\nu^{\rm eff}(z) = \int \left [ p_{\rm MS} \times \langle s_\nu^{\rm MS} \rangle + p_{\rm SB} \times \langle s_\nu^{\rm SB} \rangle \right]  \, d\textrm{sSFR},
\label{eq:effsed}
\end{equation}
where
\begin{equation}
\langle s_\nu^{\rm MS\, or \, SB}\rangle  = \int p_{\rm MS\, or \, SB} \left (\langle U \rangle \right|z) \times  s_\nu^{\rm MS\, or \, SB} \left (\langle U \rangle,z \right ) \times d \langle U \rangle
\end{equation}
where $\langle U \rangle$ is the intensity of the radiation field (strongly linked to dust temperature), $p \left (\langle U \rangle \right|z)$ its probability distribution (this encodes the information on the scatter of dust temperatures), and $s_\nu^{\rm MS\, or \, SB}$ the flux density received from an $L_{\rm IR} = 1\,L_\odot$ main-sequence or starburst source with a radiation field $\langle U \rangle$ at redshift z. The average SED of all MS or SB galaxies at any given redshift $\langle s_\nu^{\rm MS\, or \, SB}\rangle$ takes the scatter on $\langle U \rangle$ into account. The infrared luminosity of these effective templates is slightly larger than 1\,L$_\odot$ because of the asymmetry of sSFR distribution caused by starburst. The effective SEDs from Eq.\,\ref{eq:effsed} used in this work are available online\footnote{\url{http://irfu.cea.fr/Sap/Phocea/Page/index.php?id=537} or CDS via anonymous ftp to cdsarc.u-strasbg.fr (130.79.128.5) or via http://cdsweb.u-strasbg.fr/cgi-bin/qcat?J/A+A/}.\\

The contribution of satellite galaxies to differential emissivity is the sum of the contribution of all galaxies in subhalos of a main halo of mass $M_h$. It depends on the mass function of main halos, the mass function of subhalos in main halos of this mass, and the mean flux density of galaxies hosted by subhalos: 
\begin{equation}
\begin{split}
\frac{dj_{\nu,sub}}{d\textrm{log}(M_h)}(M_h,z)  = \frac{d^2N}{d\textrm{log}(M_h) dV} \times D_c^2 (1+z)\\
\times \int  s_\nu^{\rm eff}(z) \frac{dN}{d\textrm{log}(m_{\rm sub})} (m_{\rm sub}|M_h)  \times \frac{\textrm{SFR}_{\rm MS}(M_\star=f(m_{\rm sub}),z)}{K}\\
\times g(M_\star=f(m_{\rm sub}),z) \times \left (1-q \left ( M_\star=f \left ( m_{\rm sub} \right ),z \right ) \right ) d\textrm{log}(m_{\rm sub}).
\end{split}
\label{fig:jmsat}
\end{equation}
In this formula, we assume that the quenching of satellite galaxies only depends on their stellar mass (or subhalo mass). We also propose an alternative scenario where quenching depends on the mass of the main halo and not on the mass of the subhalo. In this scenario, satellite galaxies become quenched at the same time as the central galaxy in the same parent halo. In practice, we replace $m_{\rm sub}$ by $M_h$ in the last factor of Eq.\,\ref{fig:jmsat}, which can then be moved outside the integral. This scenario is motivated by the tendency for the fraction of quenched satellite galaxies to be higher in dense environments \citep[e.g.][]{Park2007}. This phenomenon is often called environmental quenching. The modified version of model B where this modification was performed is called model C.\\

In our computation, the flux densities are not the monochromatic flux densities at the center of the passband filters of each instrument, but are computed by taking the real filter profiles into account. Figure\,\ref{fig:jnu} illustrates the variation in differential emissivities with halo mass, redshift, and wavelength. The shape of the SEDs implies that long wavelengths have stronger emissivities at high redshift. The halo mass dominating the emissivities is always $\sim10^{12}\,M_\odot$, in agreement with previous works \citep[e.g.][]{Conroy2009,Bethermin2012a,Wang2012,Behroozi2012b}.\\

\subsection{Power spectrum}

The CIB power spectrum can be represented as the sum of three contributions \citep{Amblard2007,Viero2009,Planck_CIB,Amblard2011,Penin2012a,Shang2012}:
\begin{itemize}
\item Two-halo term: correlated anisotropies between galaxies in different halos, which dominates on scales larger than a few arcminutes. 
\item One-halo term: correlated anisotropies of galaxies inside the same main halo, which have a significant impact on scales of a few arcminutes.
\item Poisson term: non correlated Poisson anisotropies, also called shot noise, which dominate on small scales. 
\end{itemize}
The cross power spectrum of CIB $C_{\ell,\nu \nu'}$ between two frequency bands ($\nu$ and $\nu'$) is thus:
\begin{equation}
C_{\ell,\nu \nu'} = C_{\ell,\nu \nu'}^{2h}+C_{\ell,\nu \nu'}^{1h}+C_{\ell,\nu \nu'}^{\rm poi}
\end{equation}
\\

The two-halo and one-halo terms, which correspond to large- and intermediate-scale anisotropies, respectively, are computed from the mean emissivities of galaxies and are not affected by the stochasticity of the connexions between galaxies and halos. The computation of the anisotropies caused by the Poisson fluctuations of the number of galaxies in a line of sight requires no assumption regarding the link between dark matter halos and galaxies. They are deduced from the B12 model. Each term is calculated independently, but based on the same consistent model.\\

\subsubsection{Two-halo term}

The two-halo term is computed from the following formula, summing on redshift, but also over all cross-correlation between halos of various masses: 
\begin{equation}
\label{eq:2h}
\begin{split}
C_{\ell,\nu \nu'}^{2h}   = \iiint \frac{dD_c}{dz} \left ( \frac{a}{D_c} \right )^2 \left ( \frac{dj_{\nu,\textrm{c}}}{d\textrm{log}(M_h)}(z) + \frac{dj_{\nu,\textrm{sub}}}{d\textrm{log}(M_h)}(z) u(k,M_h,z) \right ) \\
\times \left ( \frac{dj_{\nu',\textrm{c}}}{d\textrm{log}(M'_h)}(z) + \frac{dj_{\nu',\textrm{sub}}}{d\textrm{log}(M'_h)}(z) u(k,M'_h,z) \right ) b(M_h,z) b(M'_h,z)\\
\times P_{\rm lin}(k = \frac{l}{D_c},z) \, d\textrm{log}M_h d\textrm{log}M'_h dz.
\end{split}
\end{equation}
This formula assumes the \citet{Limber1953} flat-sky approximation. The first factor is geometrical. The next two factors contain differential emissivities. The factor $u(k,M_h,z)$ is the Fourier transform of the halo profile assumed here to be NFW \citep{Navarro1997}. In contrast to \citet{Shang2012}, this term is only placed in front the subhalo-emissivity term, assuming that subhalos are distributed following the NFW profile. The central galaxy is assumed to be at the center of the halo. The clustering term $\left ( b(M_h) b(M_h') P_{\rm lin} \right ) $ is the cross power spectrum between halo of mass $M_h$ and $M_h'$, under the assumption that $P(k,M_h,M_h') = b(M_h) b(M_h') P_{\rm lin}$ \citep{Cooray2002}. Here, $P_{\rm lin}$ is the linear matter power spectrum computed with the transfer function of \citet{Eisenstein1999}. Equation\,\ref{eq:2h} can be significantly simplified by introducing
\begin{equation}
	J_\nu(z,k) = \int b(M,z) \left ( \frac{dj_{\nu,\textrm{c}}}{d\textrm{log}(M_h)} + \frac{dj_{\nu,\textrm{sub}}}{d\textrm{log}(M_h)} u(k,M_h,z) \right )dM_h, 
\end{equation}
which is an emissivity weighted by the bias corresponding to each halo mass. We can then simplify Eq.\,\ref{eq:2h}:
\begin{equation}
	\label{eq:2hsimple}
	C_{\ell,\nu \nu'}^{2h}   = \int \frac{dD_c}{dz} \left ( \frac{a}{D_c} \right )^2 J_\nu(z,k) J_\nu'(z,k) P_{\rm lin}(k = \frac{l}{D_c},z) \, dz
\end{equation}
This way of computing $C_{\ell,\nu \nu'}^{2h}$ reduces the number of integrals, because $J_\nu$ can be calculated only once per frequency channel and then can be used to derive all the cross-spectra. In addition, $J_\nu$ can also be used to compute the cross-correlation between CIB and CMB lensing (see Sect.\,\ref{sect:formciblensing}).\\

\subsubsection{One-halo term}

The one-halo term is computed with
\begin{equation}
\begin{split}
C_{\ell,\nu \nu'}^{1h} = \iint \frac{dD_c}{dz} \left ( \frac{a}{D_c} \right )^2  \times [ \frac{dj_{\nu,\textrm{c}}}{d\textrm{log}(M_h)} \frac{dj_{\nu',\textrm{sub}}}{d\textrm{log}(M_h)} u(k, M_h,z) +\\
\frac{dj_{\nu,\textrm{sub}}}{d\textrm{log}(M_h)} \frac{dj_{\nu',\textrm{c}}}{d\textrm{log}(M_h)} u(k, M_h,z) + \frac{dj_{\nu,\textrm{sub}}}{d\textrm{log}(M_h)} \frac{dj_{\nu',\textrm{sub}}}{d\textrm{log}(M_h)} u^2(k, M_h,z)]\\
	 \times \left ( \frac{d^2N}{d\textrm{log}(M_h)dV}\right )^{-1} dz \, d\textrm{log}M_h
\end{split}
\end{equation}
The first factor describes geometry. The second one represents the various (cross-)correlations between satellite and central galaxies. There is only a factor $u$ for (cross-)correlation between satellites and central, because the central is assumed to be at the center of the halo and the satellites follow the NFW profile, and $u^2$ for satellite-satellite combinations. Finally, we have to renormalize by the inverse of the mass function, because the two $dj/d\textrm{log}(M_h)$ factors implicitly contain two times the number of halos, when this should appear only once \citep{Cooray2002}. This notation avoid to have to renormalize by $j_\nu$ as in \citet{Shang2012}.

\subsubsection{Poisson term}

The Poisson term is independent of large-scale halo and only depends on the flux distribution of galaxies (number counts). These Poisson anisotropies for the auto power spectrum can be computed from \citep{Lagache2003}:
\begin{equation}
C_{\ell,\nu \nu}^{\rm poi} = \int_0^{S_{\nu,\textrm{cut}}} S_\nu^2 \frac{dN}{dS_\nu} dS_\nu. 
\end{equation}
Here $dN/dS_\nu$ are the differential number counts (see e.g. B12 for the computational details) and $S_{\nu,\textrm{cut}}$ is the flux cut at which sources are removed from the maps. The Poisson term of the cross-spectrum is slightly more complex to compute:
\begin{equation}
C_{\ell,\nu \nu'}^{\rm poi} = \int_0^{S_{\nu,\textrm{cut}}} \int_0^{S_{\nu',\textrm{cut}}} S_\nu S_\nu' \frac{d^2N}{dS_\nu \, dS_\nu'} dS_\nu dS_\nu', 
\end{equation}
where $d^2N/dS_\nu dS_{\nu'}$ are the multivariate counts (i.e. number of sources with a flux between $S_\nu$ and $S_\nu+dS_\nu$ in one band and $S_\nu'$ and $S_\nu'+dS_\nu'$ in the other).\\

In practice, multivariate counts are hard to compute, and summing the contribution of various redshift, types of galaxies, infrared luminosity, and radiation field is easier (the derivation of this formula is presented in Appendix\,\ref{sect:compcp}):
\begin{equation}
\begin{split}
C_{\ell,\nu \nu'}^{\rm poi} =  \int_z \frac{dV}{dz} \sum_{\rm \{MS, SB\}} \int_{\langle U \rangle} p_{\rm MS\, or \, SB} \left(\langle U \rangle|z \right )  \int_{L_{\rm IR}=0}^{L_{\rm IR,cut}^{\rm MS \, or \, SB}(\langle U \rangle,z)} \\
\frac{d^2N_{\rm MS\, or \, SB}}{dL_{\rm IR} dV} L_{\rm IR}^2 s_\nu^{\rm MS\, or \, SB}(\langle U \rangle,z) \times s_{\nu'}^{\rm MS\, or \, SB}(\langle U \rangle,z)\,  dL_{\rm IR} \, d\langle U \rangle \, dz
\end{split}
\end{equation}
where $L_{\rm IR,cut}^{\rm MS \, or \, SB}(\langle U \rangle,z)$ is the infrared luminosity where the source is detected in at least one of the bands (depends on redshift, type of galaxy and radiation field), and $d^2N_{\rm MS\, or \, SB}/dL_{\rm IR} \,dV$ is the infrared luminosity function (main-sequence or starburst contribution).\\

\subsection{Cross-correlation between CIB and CMB lensing}

\label{sect:formciblensing}

In addition to cross-correlations between the CIB in various bands, we can also test the predictive power of our model for correlation between the CIB and the reconstructed gravitational potential derived from distortions of the CMB due to gravitational lensing by large scale halos at z$\sim$1-3 \citep[e.g.][]{Hanson2009}. This correlation is a direct probe of the link between the gravitational potential of dark matter halos and infrared emission from star-forming galaxies.\\

Because the CMB lensing signal is due to dark matter halo, this signal can be modeled with a two-halo term replacing the term of emissivity by a term linked to the gravitational potential (adapted from \citealt{Planck_CIBxlensing}):
\begin{equation}
C_{\ell,\phi \nu} = \int  \frac{dD_c}{dz} \left( \frac{a}{D_c} \right ) J_\nu(z,\ell) \Phi(z,k) P_{\rm lin}(k = \frac{l}{D_c},z) dz,
\end{equation}
$\Phi(z,\ell)$ is given by \citep{Challinor2005}
\begin{equation}
where \Phi(z,\ell) = \frac{3}{l^2} \Omega_{\rm M} \left (\frac{H_0}{c} \right )^2 \frac{D_c}{a} \frac{D_c^{\rm CMB} - D_c(z)}{D_c^{\rm CMB} \times D_c(z)},
\end{equation}
where $H_0$ is the Hubble constant, $\Omega_{\rm M}$ the matter density in units of the critical density, and $D_c^{\rm CMB}$ the comoving distance between CMB and us. The impact of subhalos on CMB lensing (1-halo) is quite small and can be neglected for this work.\\

\begin{figure*}
\centering
\includegraphics[width=17cm]{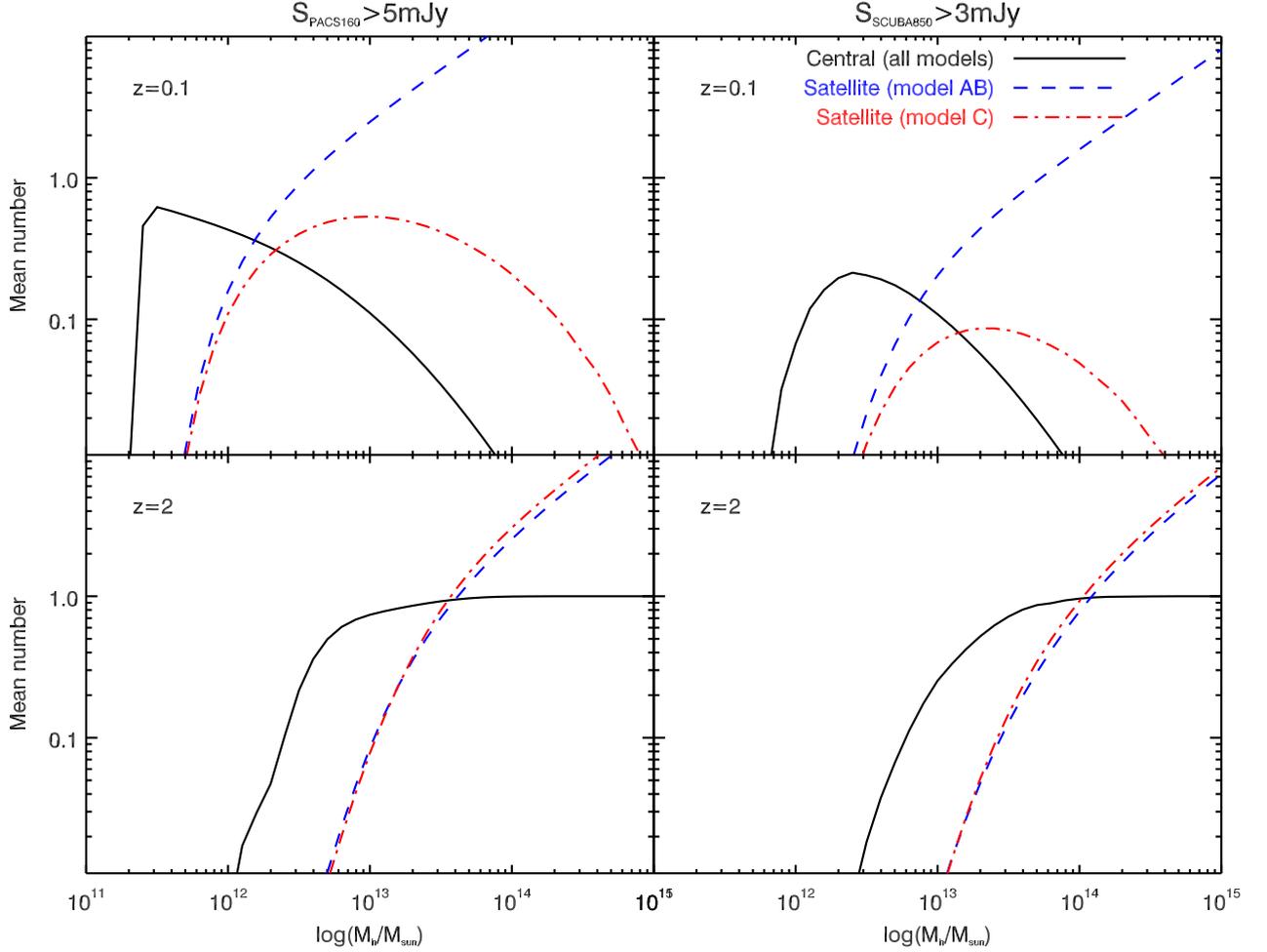} \\
\caption{\label{fig:hod} Halo occupation distribution, i.e. mean number of detected galaxies in a halo, including its substructures, as a function of its total mass $M_h$, of $S_{160}>5\,mJy$ (left) and $S_{850}>3\,mJy$ (right) sources at z = 0.1 (top) and z=2 (bottom) for central (solid line) and satellite (dashed line) galaxies predicted by the model. The HOD of central galaxies is the same for all versions of the models and represented in black. Satellite HODs are plotted in blue (red) for model A/B (C).}
\end{figure*}

\section{Computation of angular correlation function of resolved infrared galaxies}

In addition to the anisotropies of the faint infrared sources responsible for the unresolved background, our model also provides predictions of the angular correlations of the individually detected bright sources. A formalism taking the selection function of the resolved sources into account has to be employed so we cannot use the same formalism as for the power spectrum. In this paper, we only consider samples selected using a flux threshold ($S_{\nu, \textrm{cut}}$) at a given wavelength for simplicity. In this section, we first explain how we compute the halo occupation distribution (HOD, i.e. the mean number of central and satellite galaxies as a function of halo mass and redshift) for a given selection of resolved sources and then how we derive the correlation function from the HODs.

\label{sect:formalism_acf}

\subsection{Halo occupation distribution}

The mean number of central galaxies in a given halo of total mass $M_h$ is
\begin{equation}
\begin{split}
\langle N_{\rm c} \rangle (M_h,z) =   \left (1-q \left ( M_\star=f \left ( M_h \right ),z \right ) \right )\times \sum_{\rm type} \int_{\rm sSFR}  \int_{\rm \langle U \rangle}\\
H \left (s_\nu^{\rm MS \, or \, SB}(\langle U \rangle, z) \times L_{\rm IR} \left ( M_\star = f \left (M_h \right ) , \textrm{sSFR} \right ) >S_{\nu, \textrm{cut}} \right )\\
\,  d\langle U \rangle \, d {\rm sSFR},\\
\end{split}
\end{equation}
where $q \left ( M_\star=f \left ( M_h \right ),z \right )$ is the fraction of quenched objects, and $H$ is a function having a value of 1 if the condition is true and 0 otherwise. This condition is only true if the source is sufficiently bright to be detected ($S_\nu>S_{\nu, \textrm{cut}}$). This depends on halo mass, type of galaxy (MS or SB), sSFR, and $\langle U \rangle$.\\

The number of satellites depends on the total mass of the parent halo $M_h$ and is connected to the subhalo via
\begin{equation}
\langle N_{\rm sub}\rangle (M_h,z) = \int_{m_{\rm sub}} \frac{dN}{d\textrm{log}(m_{\rm sub})}(m_{\rm sub}|M_h) \langle N_{\rm c} \rangle (m_{\rm sub},z) dm_{\rm sub}.
\end{equation}
This formula works only when we assume the same infrared luminosity versus halo mass relation for main and subhalos (models A and B), and consequently the HOD of the satellite in a given subhalo of mass $m_{\rm sub}$ is the same as for the central in a main halo of identical mass $M_h = m_{\rm sub}$. For model C (satellites quenched at the same time as the centrals), the factor $\langle N_{\rm c} \rangle (m_{\rm sub},z)$ has to be multiplied by $\left (1-q \left ( M_\star=f \left ( M_h \right ),z \right ) \right )/\left (1-q \left ( M_\star=f \left ( m_{\rm sub} \right ),z \right ) \right )$ to take into account that in this version of the model the quenching is linked to the mass of the central and not the satellite.\\

Figure\,\ref{fig:hod} shows the HOD of $S_{160}>5\,mJy$ (sources detected by the deepest PACS surveys) and $S_{850}>3\,mJy$ (typical submillimeter galaxies) sources at low (z=0.1) and high (z=2) redshift. The number of central galaxies decreases very quickly at low mass. This sharp cut is due to the selection in infrared flux and the rather tight correlation between infrared luminosity and halo mass for central galaxies. At low redshift, the probability to detect an infrared galaxy in a massive halo is fairly low because the vast majority of central galaxies are quenched and thus have little infrared emission. The HOD of the satellites strongly depends on the version of the model. If the quenching of satellites is decorrelated from the quenching of the centrals (models A and B), the number of the detected satellite increases quickly with the halo mass. This is not the case for model C, for which the satellites are quenched at the same time as the centrals.\\

\subsection{Angular auto-correlation function}

The angular correlation function (ACF) can be compute from the HOD of galaxies and their redshift distribution (see \citealt{Cooray2002} for a review):
\begin{equation}
w(\theta) = \frac{\int_z \left ( \frac{dN}{dz} \right )^2 \int_{k} \frac{k}{2 \pi} P_{\rm gg}(k,z) J_0(k D_c \theta) \, dz \, dk}{\left ( \int_z \frac{dN}{dz}  \, dz \right )^2},
\end{equation}
where $J_0$ is the zeroth order Bessel function, and $dN/dz$ the redshift distribution of galaxies, which can be computed from counts per redshift slice following \citet{Bethermin2011}. The value $P_{\rm gg}$ is the sum of two terms corresponding to the clustering of galaxies in two different halos and inside the same halo:
\begin{equation}
P_{\rm gg}(k,z) = P_{\rm gg}^{\rm 2h}(k,z) + P_{\rm gg}^{\rm 1h}(k,z).
\end{equation}
Following the standard conventions, we write the two-halo term as
\begin{equation}
P_{\rm gg}^{\rm 2h}(k,z) =  \left [ \int_{M_h} \frac{d^2N}{d\textrm{log}(M_h) dV} b(M_h) \frac{\langle N_{\rm gal} \rangle}{\bar{n}_{\rm gal}} dM_h\right ]^2 P_{\rm lin}(k,z),
\end{equation}
where $\langle N_{\rm gal} \rangle = \langle N_{\rm c} \rangle  + \langle N_{\rm sub} \rangle$. The one-halo term is
\begin{equation}
\begin{split}
P_{\rm gg}^{\rm 1h}(k,z) =  \int_{M_h} \frac{d^2N}{d\textrm{log}(M_h) dV} \frac{2 \langle N_{\rm c} \rangle \langle N_{\rm sub} \rangle + \langle N_{\rm sub} \rangle^2 u^2(k,M_h,z)}{\bar{n}_{\rm gal}}\\
dM_h,
\end{split}
\end{equation}
where
\begin{equation}
\bar{n}_{\rm gal} = \int  \frac{d^2N}{d\textrm{log}(M_h) dV} \langle N_{\rm gal} \rangle dM_h.
\end{equation}

\begin{figure*}
\centering
\includegraphics[width=17.6cm]{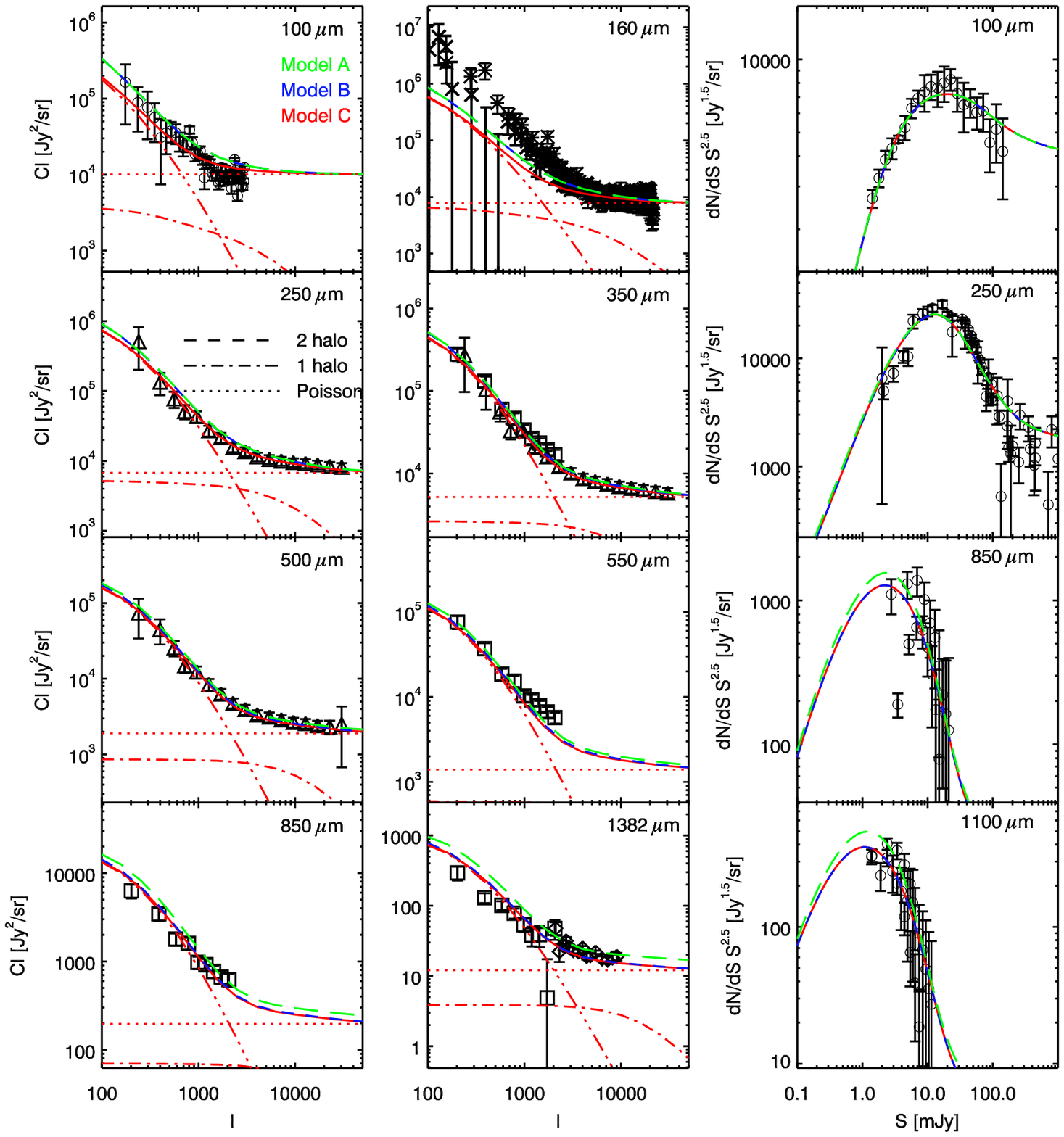}
\caption{\label{fig:autospec} \textit{Left and center panels}: CIB power spectrum predicted by our model and comparison with IRAS (160\,$\mu$m, \citealt{Penin2012b}, \textit{open circles}), \textit{Spitzer} (100\,$\mu$m, \citealt{Lagache2007}, \textit{crosses}, \citealt{Penin2012b}, \textit{asterisks}), \textit{Herschel} (250, 350, and 500\,$\mu$m, \citealt{Viero2012}, \textit{triangles}), \textit{Planck} (350, 550, 850, and 1380\,$\mu$m, \citet{Planck_CIB}, \textit{squares}), and SPT (1380\,$\mu$m, \citealt{Hall2010}, \textit{diamonds}) measurements. The \textit{dotted}, \textit{dot-dashed}, and \textit{dotted} lines represent the Poisson, 1-halo, and 2-halo terms. \textit{Right panel}: Number counts of infrared galaxies. Data are taken from the compilation of measurements in \citet{Bethermin2011} and B12. Models A, B, and C are represented by a long green dash, a short blue dash, and a solid red line, respectively. The flux cuts used to compute the model predictions are 1\,Jy at 100\,$\mu$m, 100\,mJy at 160\,$\mu$m, 0.3\,mJy at 250\,$\mu$m, 350\,$\mu$m, and 500\,$\mu$m, 0.54\,Jy at 550\,$\mu$m, 0.325\,Jy at 850\,$\mu$m, 20\,mJy at 1.38\,mm.} 
\end{figure*}

\begin{figure*}
\centering
\includegraphics[width=17.5cm]{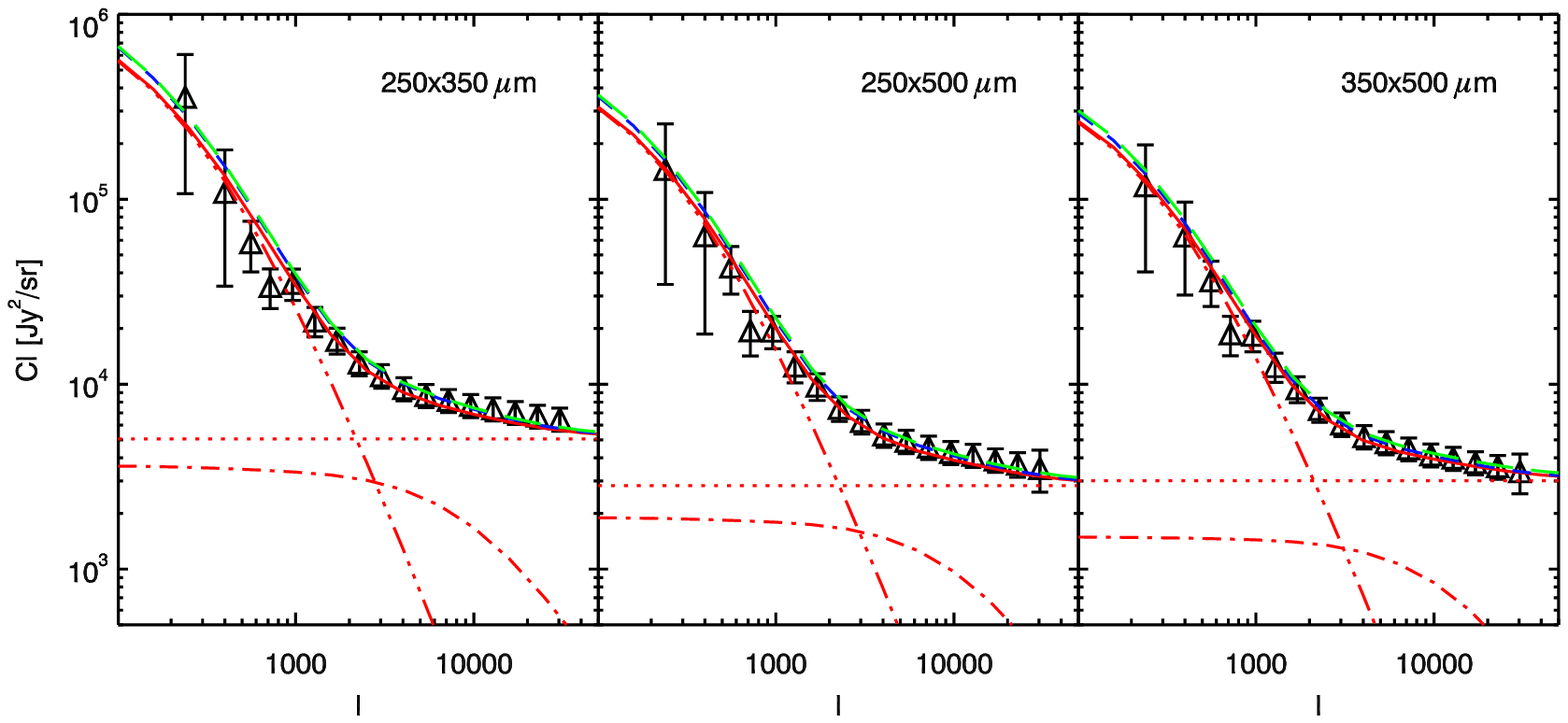}
\caption{\label{fig:xspire} Cross-power spectrum between SPIRE bands measured by \citet{Viero2012} and comparison with our model. Models A, B, and C are represented by a long green dash, a short blue dash, and a solid red line, respectively.}
\end{figure*}

\begin{figure*}
\centering
\includegraphics[width=17.5cm]{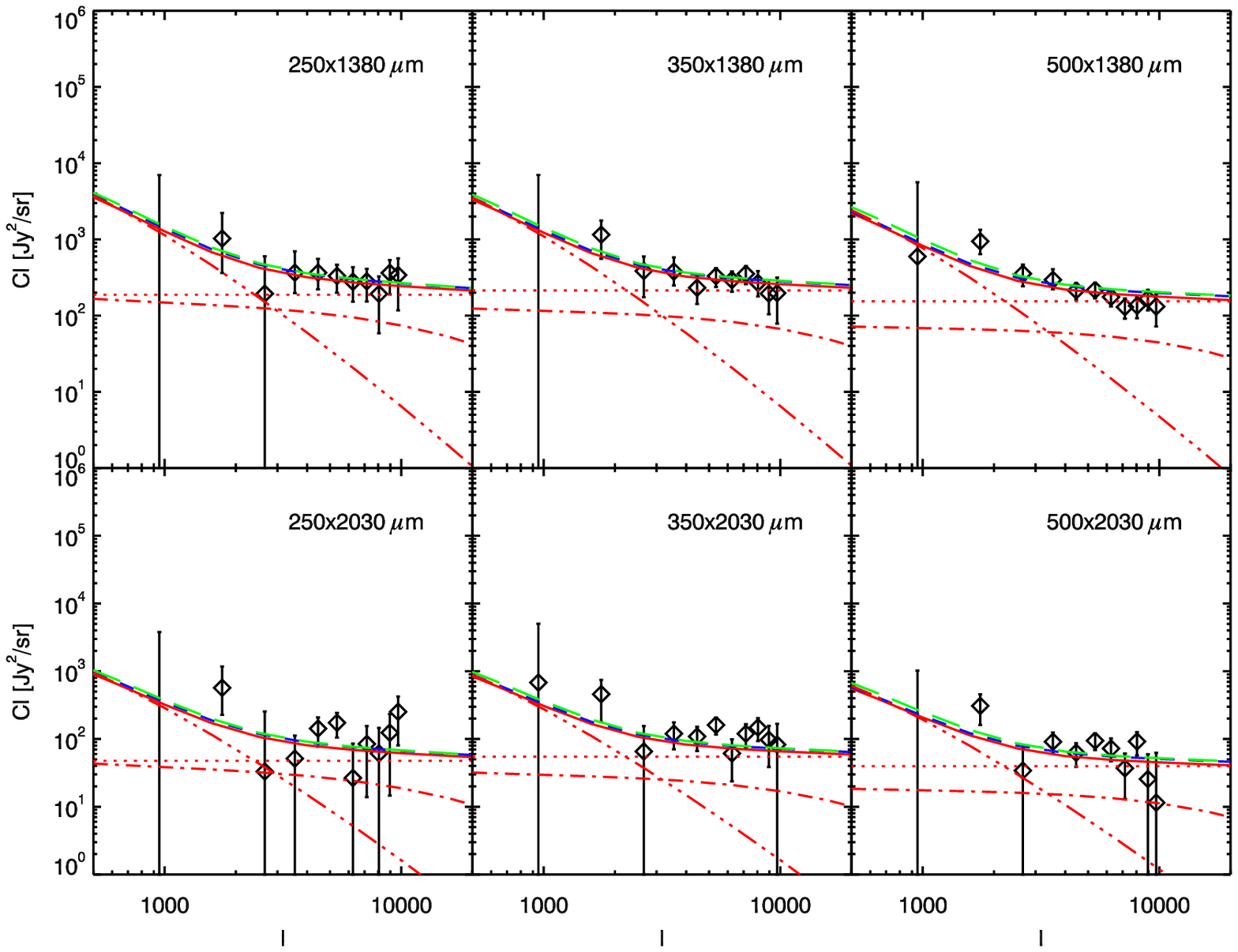}
\caption{\label{fig:blastxact} Cross-power spectrum between BLAST and ACT measured by \citet{Hajian2011} and comparison with our model. Models A, B, and C are represented by a long green dash, a short blue dash, and a solid red line, respectively.}
\end{figure*}

\begin{figure}
\centering
\includegraphics[width=7.5cm]{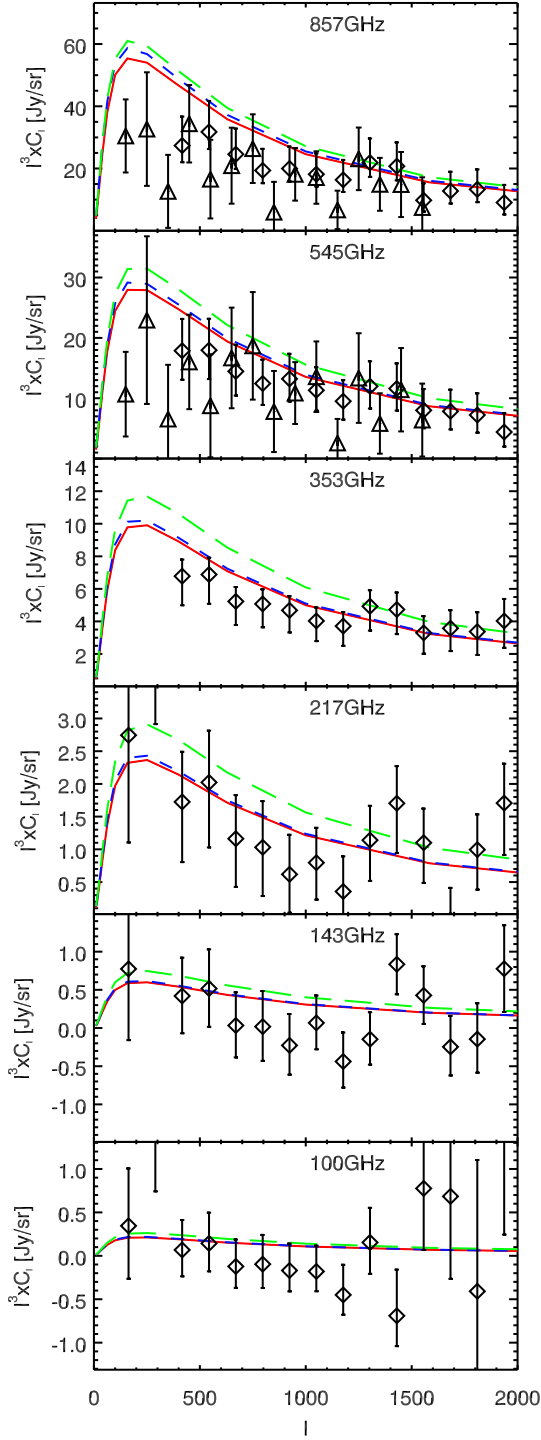}
\caption{\label{fig:cibxlensing} Comparison between the cross power spectrum between CIB and CMB lensing predicted by our model and the measurements by \citet{Planck_CIBxlensing} (diamonds) and \citet{Holder2013} with SPT and \textit{Herschel} (triangles). Models A, B, and C are represented by a long green dash, a short blue dash, and a solid red line, respectively.}
\end{figure}

\begin{figure}
\centering
\includegraphics[width=7.5cm]{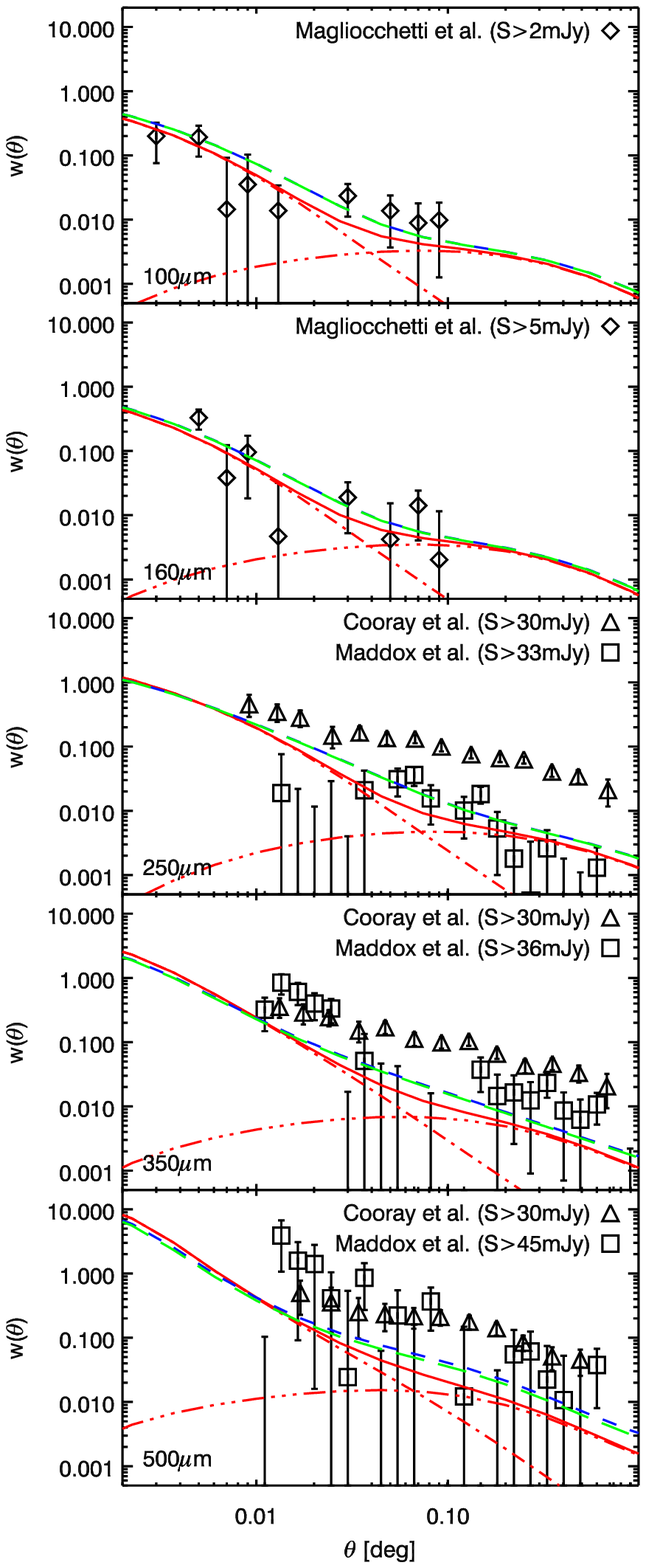}
\caption{\label{fig:wtheta} Auto-correlation function of various flux-selected sample of infrared galaxies and comparison with our model (same color coding as in Fig.\,\ref{fig:autospec}). The data come from \citet{Magliocchetti2011} at 100$\mu$m and 160\,$\mu$m and \citet{Cooray2010} and \citet{Maddox2010} at 250, 350, and 500\,$\mu$m. We used the same flux cuts as \citet{Magliocchetti2011} and \citet{Cooray2010}. Models A, B, and C are represented by a long green dash, a short blue dash, and a solid red line, respectively.}
\end{figure}

\section{Comparison with the observations}

\label{sect:results}

\subsection{CIB power spectrum and number counts}

We compared the predictions of our model with the CIB power spectrum measured with IRAS (100\,$\mu$m, \citealt{Penin2012b}), \textit{Spitzer} (160\,$\mu$m, \citealt{Lagache2007,Penin2012b}), \textit{Herschel} (250, 350, and 500\,$\mu$m, \citealt{Viero2012}), \textit{Planck} (350, 550, 850, and 1380\,$\mu$m, \citet{Planck_CIB}), and SPT (1380\,$\mu$m, \citealt{Hall2010}) measurements in Fig.\,\ref{fig:autospec} (left and center). The level of the Poisson anisotropies is significantly affected by the flux cut used to mask bright sources, and their level increases with the chosen flux cut. The correlated anisotropies (1-halo and 2-halo) are less affected by the flux cut. When we compare several datasets at the same wavelength, we thus compute the model predictions for the flux cut of the experiment, which is the most sensitive on small scales. The flux cuts used to compute the model predictions are thus 1\,Jy at 100\,$\mu$m, 100\,mJy at 160\,$\mu$m, 0.3\,mJy at 250\,$\mu$m, 350\,$\mu$m, and 500\,$\mu$m, 0.54\,Jy at 550\,$\mu$m (545\,GHz), 0.325\,Jy at 850\,$\mu$m (353\,GHz), 20\,mJy at 1.38\,mm (217\, GHz). The overall agreement with the data is very good for a model not fitted on the data. However, some tensions between models A and B and data at short and large wavelengths provide interesting information.\\

The anisotropies at 100\,$\mu$m are dominated by galaxies at low redshift (see Sect.\,\ref{sect:origin}) and are thus not affected by the evolution of galaxies at z$>$2.5. Consequently, there is no difference between models A and B. In contrast, model C (with quenching of satellite around massive quenched central) predicts a lower level of anisotropies at $\ell<5000$, which is in better agreement with the data, especially around $\ell = 2000$. The "environmental quenching" in model C reduces the one-halo term compared to model B (by a factor of 4 at $\ell$=2000), but does not  significantly affect the two-halo term ($\sim$15\%) and does not change the Poisson level at all. However, the former difference is not very significant ($\sim 2\sigma$), and CIB anisotropies are obviously not the best probe of such environmental effects.\\

The interpretation of \textit{Spitzer}/MIPS 160\,$\mu$m data is trickier. While our model agrees with the measurements of \citet{Lagache2007} (crosses) and  \citet{Penin2012b} (asterisks) at $\ell>2000$, they are a factor 3-10 higher than our model on larger scales. Such a large difference is hard to explain especially when considering that the model agrees well with the data at 100 and 250\,$\mu$m. This could be caused by problems in cirrus subtraction. In addition, MIPS suffers from strong 1/f noise and is probably not the best instrument for measuring large-scale anisotropies.\\

Between 250 and 550\,$\mu$m, our model agrees nicely with \textit{Planck} and \textit{Herschel} measurements. At longer wavelengths (850\,$\mu$m and 1.4\,mm), model A overpredicts anisotropies on large scales by a factor 2 at 1.4\,mm, while models B and C agree with the data, except for a 2$\sigma$ tension at 1.4\,mm in the two lowest multipole bins. However, measuring the CIB at 217\,GHz is difficult because it strongly relies on the correct subtraction of the CMB. Future analyses will either confirm or refute the presence of this discrepancy. The three models agree with the number counts at 850\,$\mu$m and 1.1\,mm. The difference between models A and B/C is the sSFR at z$>$2.5 and the characteristic density of the mass function of star-forming galaxies. Model A assumes a flat sSFR at z$>$2.5, while models B and C assume rising sSFR at z$>$2.5 compensated for by a decrease in the characteristic density of the mass function to preserve the agreement with the number counts. In these two versions, the number of bright objects is thus the same, but the number of faint objects is different (see e.g. number counts around 1\,mJy at 1.1\,mm). Model A overpredicts the CIB anisotropies because of too high emissivity of high-redshift faint galaxies.\\
 

In the case of model C (which includes satellite-quenching), the one-halo term never dominates the anisotropies regardless of wavelength or scale. Concerning models A and B (without satellite-quenching), the one-halo term contributes to the CIB anisotropies by a similar amount as do the two-halo and Poisson terms at $\ell\sim2000$ at 100\,$\mu$m, 160\,$\mu$m, and 250\,$\mu$m. On other scales and wavelengths, it represents a minor contribution. This result contradicts models assuming no mass-luminosity dependence \citep[e.g.][]{Planck_CIB,Penin2012a}, but is consistent with the approaches assuming a dependence \citep[e.g.][]{Shang2012}. This reduction of the one-halo term is caused by the satellites being hosted by low mass subhalos and thus having low infrared luminosities.\\

\subsection{CIB cross-power spectrum}

In addition to auto power spectra, we can compare the predictions of our model with cross power spectra. They are useful for verifying that the level of correlation between bands is correct. This also is an indirect test of whether SED libraries and redshift distributions used in the model are correct. We compared our models with measurements of \citet{Viero2012} between SPIRE bands (see Fig.\,\ref{fig:xspire}). All versions of our model agree with the data, except a systematic 2$\sigma$ tension for the point at $\ell=700$ in the three bands, which could be due to cosmic variance (one 2$\sigma$ outlier for 18 points is statistically expected). The trends of \citet{Hajian2011} (see Fig.\,\ref{fig:blastxact}) are also reproduced well, but our model is systematically lower than the data at $\ell=1500$. However, the stability of BLAST on large scales is not as good as for SPIRE. A future analysis of the cross-correlation between SPIRE and ACT and/or between \textit{Planck} bands will thus be useful for further investigating of this discrepancy.\\ 

\subsection{Cross correlation between CIB and CMB lensing}

In Fig.\,\ref{fig:cibxlensing}, we compare the predictions of the cross-correlation between CIB and CMB lensing with the measurements from \citet{Planck_CIBxlensing} and \citet{Holder2013}. In the former work, both CMB lensing and CIB fluctuations are measured from \textit{Planck} data. In the latter, the CMB lensing is estimated from SPT data, and the CIB is measured using \textit{Herschel}. At $\ell>$1000, all versions of the model agree well with the data. At $\ell<$1000, the various versions of the model tend to be be systematically higher than the data. Models B and C (with modified evolution at high redshift) are closer to the data, but slightly overpredict the cross-correlation by $\sim$1$\sigma$ at 857\,GHz, 545\,GHz, and 353\,GHz. Future work will investigate this tension.\\

\subsection{Clustering of resolved galaxies}

The clustering of bright resolved infrared sources is also an important test of our model. Figure\,\ref{fig:wtheta} shows the comparison between the clustering  measurements of various flux-selected samples and our model predictions. At 100 and 160\,$\mu$m, our model agrees well with the \textit{Herschel}/PACS measurements of \citet{Magliocchetti2011}. In \textit{Herschel}/SPIRE bands, the comparison is trickier. There are strong disagreements between the measurements of \citet{Cooray2010} and \citet{Maddox2010}. In fact, measuring the correlation function of SPIRE sources (250, 350, and 500\,$\mu$m) is very difficult because these data are strongly confusion-limited \citep{Nguyen2010}. The background is thus hard to estimate, and the completeness of the catalog can vary depending on the local source density. We agree with \citet{Maddox2010} on scales larger than 0.1\,deg, but not on smaller scales at 250\,$\mu$m and 350\,$\mu$m, where systematics could be due to background subtraction. Our model disagrees with the measurements of \citet{Cooray2010}, which are systematically higher. More reliable measurements of the correlation function of SPIRE sources, controlling the systematic effects, are thus needed to check the validity of our model accurately.

\section{Successes and limitations of our model}

\label{sect:validity}

In the previous sections, we presented a natural way of extending the B12 model of infrared galaxies by linking their properties to their host halo. We tested the validity of this approach by comparing the prediction of this extended model to the measured spatial distribution of both individually detected infrared galaxies (where the measurements are reliable) and the unresolved background. This comparison shows the good predictive power of our approach, suggesting that our assumptions are fair. Using this model to interpret both the origin of CIB anisotropies (Sect.\,\ref{sect:origin}) and the link between star-forming galaxies and dark matter halos (Sect.\,\ref{sect:sfe}) is thus legitimate. For simplicity, we perform this analyses only with model C, which provides the best agreement with the data.\\

Few models are able to reproduce both the number counts from mid-infrared to the millimeter wavelengths and CIB anisotropies from the far-infrared to the millimeter regimes. \citet{Addison2013} proposed a fully empirical model based on the evolution of the luminosity function and the bias of galaxies. This approach is as efficient as ours when it comes to fitting counts, but provides very little insight into the physical evolution of galaxies. \citet{Cai2013} -- an updated and extended version of models presented in \citet{Lapi2011} and \citet{Xia2012} models -- is also very efficient at reproducing all these observables, despite too low a level of the CIB anisotropies below 350\,$\mu$m. However, their approach has many free parameters (32 for the evolution of the luminosity function of the various low-z galaxy populations, ten for the physically-modeled population of high-z proto-spheroidal galaxies, and four for the clustering of galaxies), and uses a description of the clustering based on an HOD standard that is not coupled to the physical model describing the evolution of their protospheroidal galaxies. The strength of our approach is to propose a set of 18 parameters for the evolution of galaxies, which are all constrained using external constraints (measurements of the evolution of the sSFR and the SMF, mean SEDs measured by stacking, etc.), and a formalism to compute the clustering without any free parameters, allowing a consistent description of the occupation of dark-matter halos by infrared galaxies.\\

However, some limitations of this model have to be kept in mind. First, the SED of infrared galaxies at high redshift ($z>2$) was not measured and it is not clear if the dust temperature will increase or decrease at high redshift, because of the uncertainties on the mass-metallicity and mass-attenuation relation as discussed in \citet{Magdis2012}. The good agreement between the model and the CIB anisotropies at long wavelengths suggests that our hypothesis of nonevolution is reasonable. The SED of low-mass galaxies ($M_\star<10^{10}\,M_\odot$) is not well constrained, but the contribution of these low-mass galaxies to the CIB is small making this scenario hard to test. Second, the specific star formation rate and the SMF of star-forming galaxies at very high redshift ($z>3$) are also uncertain, but the effects of $z>3$ galaxies on the CIB are pretty small, and these data thus cannot accurately constrain the evolution of infrared galaxies at very high redshift. Finally, we assumed an universal relation between stellar mass and UV attenuation, but this relation could also break at very high redshift, where the amount of metals is smaller and galaxies could thus contain less dust. For all these reasons, the predictions of the models at $z>2$ must be interpreted with caution.

\begin{figure*}
\centering
\includegraphics[width=15.8cm]{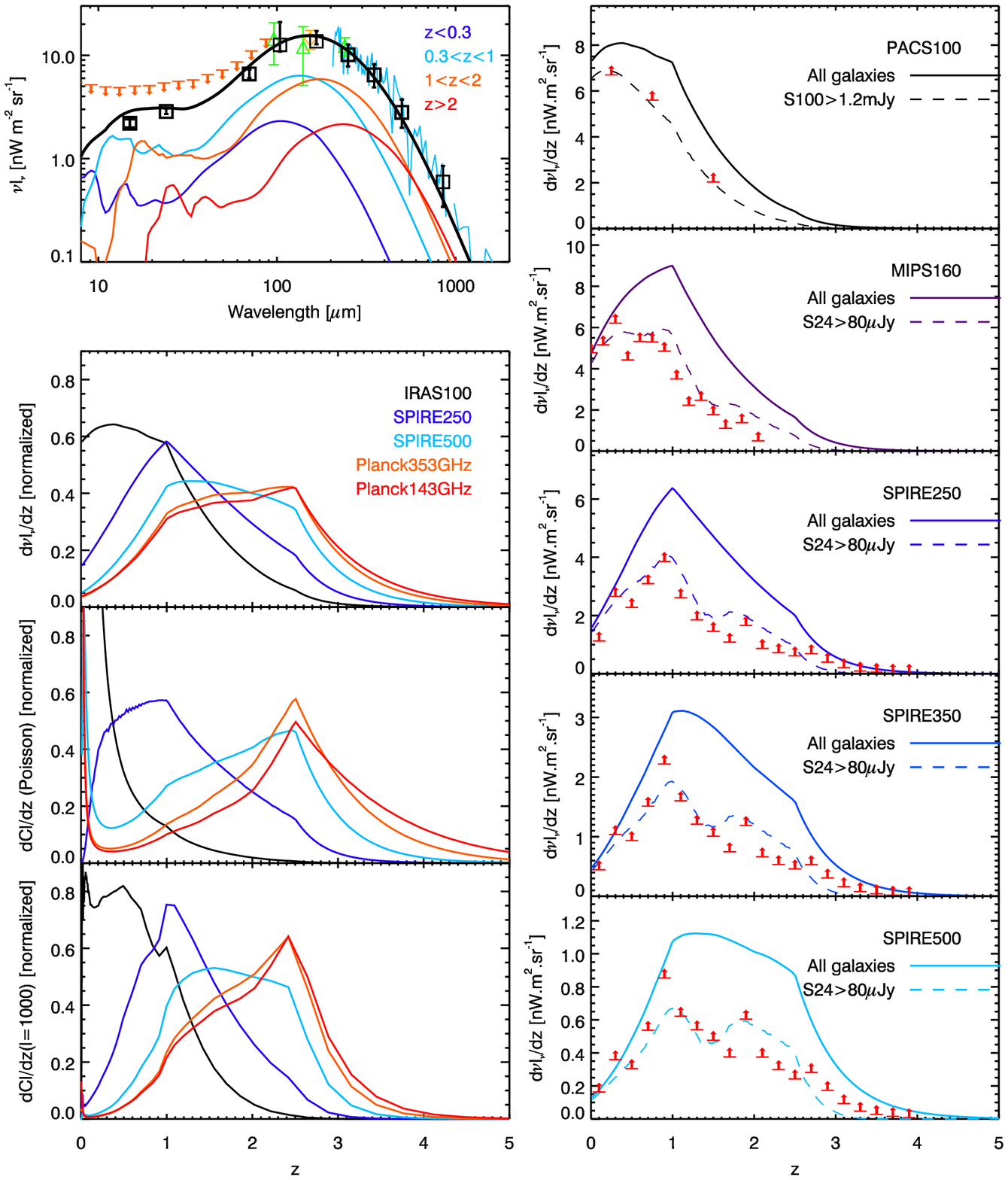}
\caption{\label{fig:dcibdz} Predicted redshift distribution of the CIB and its anisotropies. \textit{Upper left panel:} CIB SED and contribution per redshift slice (colored solid lines). \textit{black squares}: total extrapolated CIB from deep number counts \citep{Teplitz2011,Bethermin2010a,Berta2011,Bethermin2012b,Zemcov2010}. \textit{Cyan solid line:} Absolute CIB spectrum measured by \textit{COBE}/FIRAS \citep{Lagache2000}. \textit{Green triangles}: absolute CIB measurements performed by \textit{COBE}/DIRBE at 100\,$\mu$m, 140\,$\mu$m, and 240\,$\mu$m (updated in \citealt{Dole2006}). \textit{Yellow diamond:} absolute measurements of \citet{Penin2012a} at 160\,$\mu$m with \textit{Spitzer}/MIPS. \textit{Orange arrows}: upper limits derived from opacity of the Universe to TeV photons \citep{Mazin2007}. \textit{Lower left panel}: Normalized redshift distribution of the mean level (up), Poisson anisotropies (middle), large scale anisotropies at $\ell$=1000 (bottom). Various colors correspond to various bands. \textit{Right panel}: Contribution of various redshifts to the CIB (solid line) and comparison with lower limits derived by stacking  from \citet{Berta2011} at 100\,$\mu$m, \citet{Jauzac2011} at 160\,$\mu$m and \citet{Bethermin2012b} at 250\,$\mu$m, 350\,$\mu$m, and 500\,$\mu$m (arrows). The dashed lines are the model predictions taking the selection used to derive the lower limits into account. The flux density cuts are the same as previously (see Sect.\,\ref{sect:results} and Fig.\,\ref{fig:autospec}).}
\end{figure*}

\begin{figure*}
\centering
\includegraphics[width=17cm]{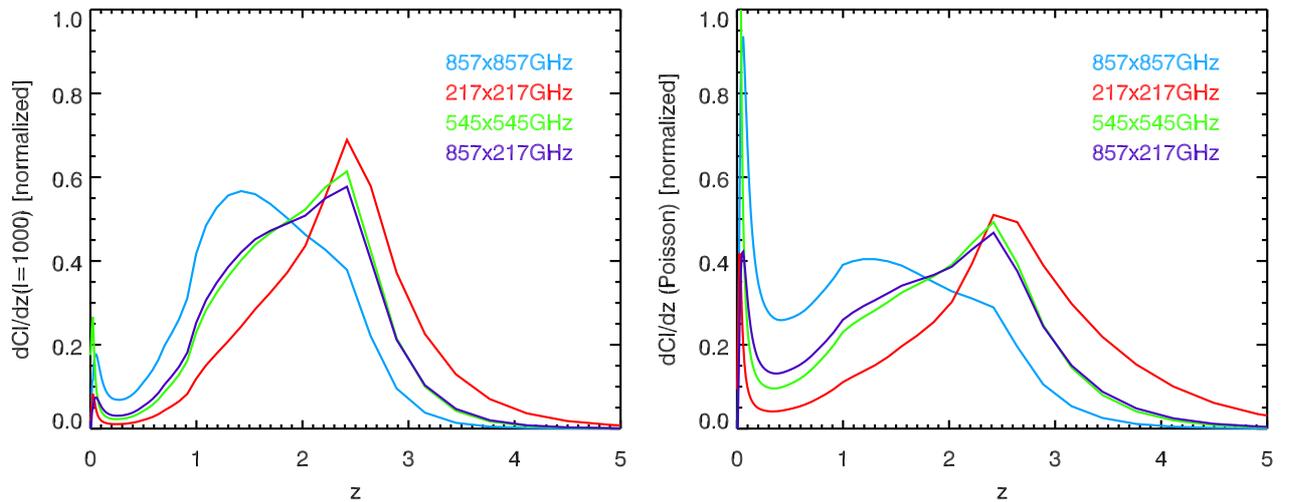}
\caption{\label{fig:dbdz_xspec} Predicted redshift distribution of auto and cross power spectra at large ($\ell=1000$, left) and small scale (Poisson level). We focus on three frequencies: 857\,GHz (350\,$\mu$m), 545\,GHz (550\,$\mu$m), and 217\,GHz (1382\,$\mu$m) to illustrate the difference between cross spectrum between two distant wavelengths and the auto-spectrum at an intermediate wavelength. The flux density cuts are the same as previously (see Sect.\,\ref{sect:results} and Fig.\,\ref{fig:autospec}).}
\end{figure*}

\begin{table*}
\centering
\caption{\label{tab:correl} Predicted correlation ($C_\ell^{\nu \nu'} / \sqrt{C_\ell^{\nu \nu} \times C_\ell^{\nu' \nu'}}$) of CIB anisotropies between bands on small (Poisson level, upper part) and large scales (l=1000, lower part).}
\begin{tabular}{lrrrrrrr}
\hline
\hline
\multicolumn{8}{c}{Correlation between bands on small scales (Poisson)} \\
\hline
 & 3000\,GHz & 857\,GHz & 545\,GHz& 353\,GHz & 217\,GHz & 143\,GHz & 100\,GHz \\
\hline
3000\,GHz & 1.000 &  &  &  &  &  &  \\
857\,GHz & 0.599 & 1.000 &  &  &  &  &  \\
545\,GHz & 0.407 & 0.916 & 1.000 &  &  &  &  \\
353\,GHz & 0.310 & 0.830 & 0.962 & 1.000 &  &  &  \\
217\,GHz & 0.277 & 0.785 & 0.920 & 0.968 & 1.000 &  &  \\
143\,GHz & 0.279 & 0.745 & 0.894 & 0.963 & 0.958 & 1.000 &  \\
100\,GHz & 0.333 & 0.743 & 0.881 & 0.940 & 0.921 & 0.988 & 1.000 \\
\hline
\hline
\multicolumn{8}{c}{Correlation between bands on large scale (l=1000)} \\
\hline
 & 3000\,GHz & 857\,GHz & 545\,GHz & 353\,GHz & 217\,GHz & 143\,GHz & 100\,GHz \\
\hline
3000\,GHz & 1.000 &  &  &  &  &  &  \\
857\,GHz & 0.766 & 1.000 &  &  &  &  & \\
545\,GHz & 0.636 & 0.971 & 1.000 &  &  &  &  \\
353\,GHz & 0.555 & 0.925 & 0.988 & 1.000 &  &  &  \\
217\,GHz & 0.529 & 0.902 & 0.975 & 0.997 & 1.000 &  & \\
143\,GHz & 0.538 & 0.904 & 0.975 & 0.996 & 0.999 & 1.000 &  \\
100\,GHz & 0.575 & 0.919 & 0.981 & 0.996 & 0.997 & 0.999 & 1.000 \\

\hline
\end{tabular}
\end{table*}

\begin{figure}
\centering
\includegraphics[width=8cm]{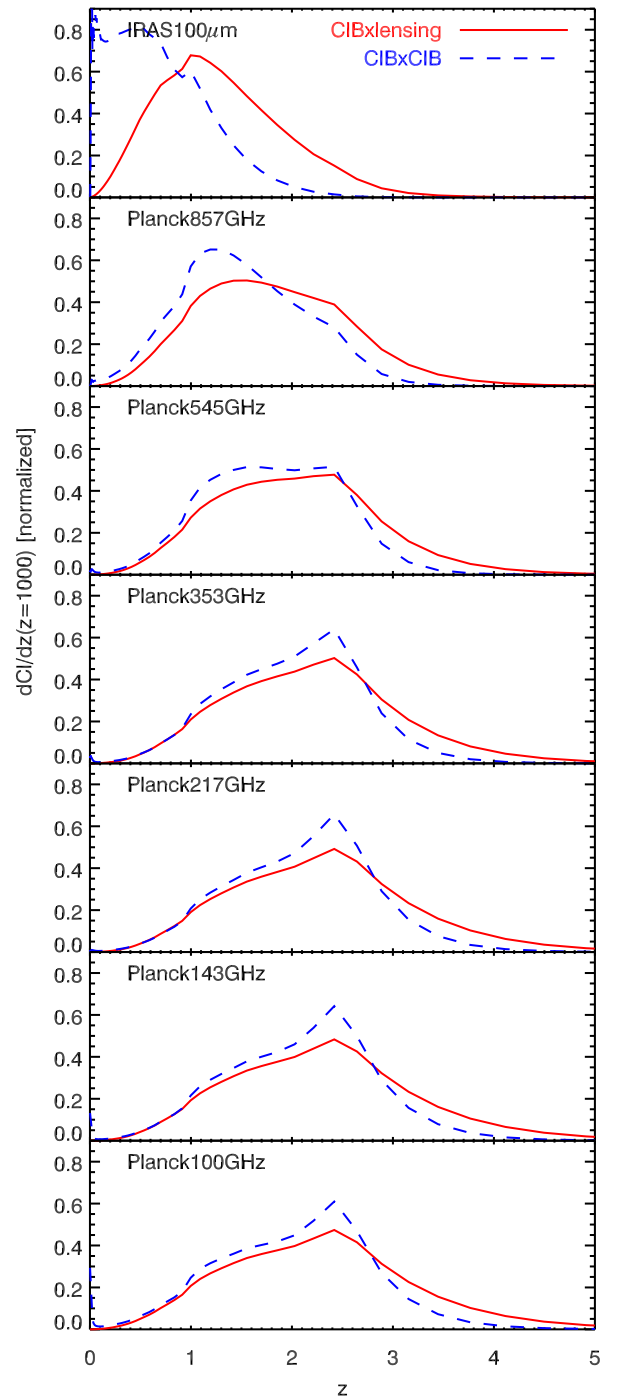}
\caption{\label{fig:Nz_lens} Comparison between the predicted redshift distributions of CIBxCIB and CIBxCMB lensing signal in various IRAS/\textit{Planck} bands. \textit{Solid red line}: Redshift distribution of the cross-correlation between CIB and CMB lensing at l=1000 in various bands. \textit{Dashed blue line}: Redshift distribution of CIB auto power spectrum.}
\end{figure}

\section{Where do the CIB and its anisotropies come from?}

\label{sect:origin}

In this section, we discuss the predictions of the model concerning the redshift distribution of the various signals we studied (CIB auto-spectra, CIB cross-spectra, CIBxCMB lensing). We base our analysis on the model C as justified in Sect.\,\ref{sect:validity}.

\subsection{Redshift distribution of CIB mean level and anisotropies}

The CIB SED can be predicted directly from the \citet{Bethermin2012c} count model. Figure\,\ref{fig:dcibdz} (upper left panel) shows the spectral energy distribution of the CIB and its decomposition per redshift slice. Compared to the \citet{Bethermin2011} model, this new model has a slightly higher contribution of $1<z<2$ sources. This new model agrees well with both absolute measurements and the total contribution of infrared galaxies extrapolated from the number counts.\\

The three lower lefthand panels of Fig.\,\ref{fig:dcibdz} show the redshift distribution of the intensity of the CIB (top), but also of its anisotropies on small (middle) and large scales (bottom). To allow an easier comparison between bands, we normalized the redshift distributions in order to have $\int d(\nu I_\nu)/dz\, dz = 1$ and $\int dC_\ell/dz\, dz = 1$. Between 100\,$\mu$m (3000\,GHz) and 850\,$\mu$m (353\,GHz), the redshift distribution evolves strongly toward higher redshift. At longer wavelengths, there is almost no evolution, because all the sources below $z\sim7$ are seen in the Rayleigh-Jeans regime and the color is roughly the same at all redshifts. This trend is seen for both intensity and anisotropies, regardless of scale. There are, however, small differences between these three quantities. For instance, the redshift distribution of the Poisson level\footnote{We used the same flux cuts as in Sect.\,\ref{sect:results}} at 100\,$\mu$m is dominated by $z<0.5$ sources, when a significant fraction of $\ell=1000$ anisotropies is caused by galaxies at z$\sim$1. This is because the Poisson anisotropies are dominated by a small number of low-z bright galaxies just below the flux cut. Anisotropies on large scales are dominated by normal star-forming galaxies at z$\sim$1 which dominate the background and the two-halo term of anisotropies.\\

Finally, we checked that the redshift distributions predicted by our model agree with the lower limits derived by stacking of 24\,$\mu$m sources. We compare them with the lower limits from \citet{Berta2011} at 100\,$\mu$m, \citet{Jauzac2011} at 160\,$\mu$m and \citet{Bethermin2012b} at 250\,$\mu$m, 350\,$\mu$m, and 500\,$\mu$m, which agree well with our model. We used our model to simulate the selection corresponding to these various works and found good overall agreement below z$\sim$2.5. At higher redshift and for a 24\,$\mu$m selection, these predictions are underestimated. This is expected because our SED templates only take the dust emission into account, when the $\lambda<8$\,$\mu$m rest-frame emission is significantly affected by stellar emissions. We thus underestimate the number of 24\,$\mu$m detections and consequently also their contribution to the CIB.\\

\subsection{Redshift distribution of CIB cross-correlation between bands}

Cross power spectra between bands provide complementary information to the auto-spectra. Figure\,\ref{fig:dbdz_xspec} shows the redshift distribution of a selection of cross and auto-spectra. On both large ($\ell=1000$, left) and small scales (Poisson), the redshift distribution of the cross-spectra between two distant bands (e.g., 857 and 217\,GHz here) is fairly close to the redshift distribution for an intermediate band (e.g., 545\,GHz here), except for a difference in Poisson level at very low z caused by different flux cuts. Considering the quite fine spectral coverage of our data, we thus cannot claim that cross-spectra probe different redshift ranges.\\

In contrast, the correlation ($C_\ell^{\nu \nu'} / \sqrt{C_\ell^{\nu \nu} \times C_\ell^{\nu' \nu'}}$) between bands provides a useful test of the validity of the models (especially their SEDs and redshift distributions). This correlation varies with scale. We thus focused on Poisson level and $\ell=1000$. The results are summarized in Table\,\ref{tab:correl}. The level of correlation is also important for cosmic microwave background (CMB) experiments to know whether the CIB at high frequency is a good proxy for the CIB emission at low frequency. For instance, the correlation between 857\,GHz and 143\,GHz is 0.904 at $\ell=1000$, but only 0.75 for the Poisson level. This correlation is predicted to rise to 0.996 between 353\,GHz and 143\,GHz (0.963 for the Poisson level). According to our model, the CIB at 353\,GHz thus provides a very good proxy for CIB in the CMB channels.\\

\subsection{Redshift distribution of CIBxCMB lensing signal}

The cross-correlation between CIB and lensing provides supplementary information probing a slightly different redshift range. Figure\,\ref{fig:Nz_lens} shows a comparison between the predicted redshift distributions of the CIB auto-spectrum and the cross-spectrum between CIB and CMB lensing for various bands. While the redshift distribution of the auto-spectrum evolves strongly between 100 (3000\,GHz) and 850\,$\mu$m (353\,GHz), this evolution is weaker for the cross-correlation with CMB lensing that mainly probes the $z=1-2$ redshift range. CIBxCMB lensing is thus a good probe of the mean SED of galaxies contributing to the CIB at this epoch. However, apart from the 100\,$\mu$m band, the redshift range probed by CIBxCMB lensing is relatively similar to that of the CIB power spectrum, but it is an independent and more direct probe of the link between dark matter and star-forming objects, because the lensing of the CMB by the large scale structures is a very well known and modeled physical process.\\

\begin{figure*}
\centering
\includegraphics[width=17cm]{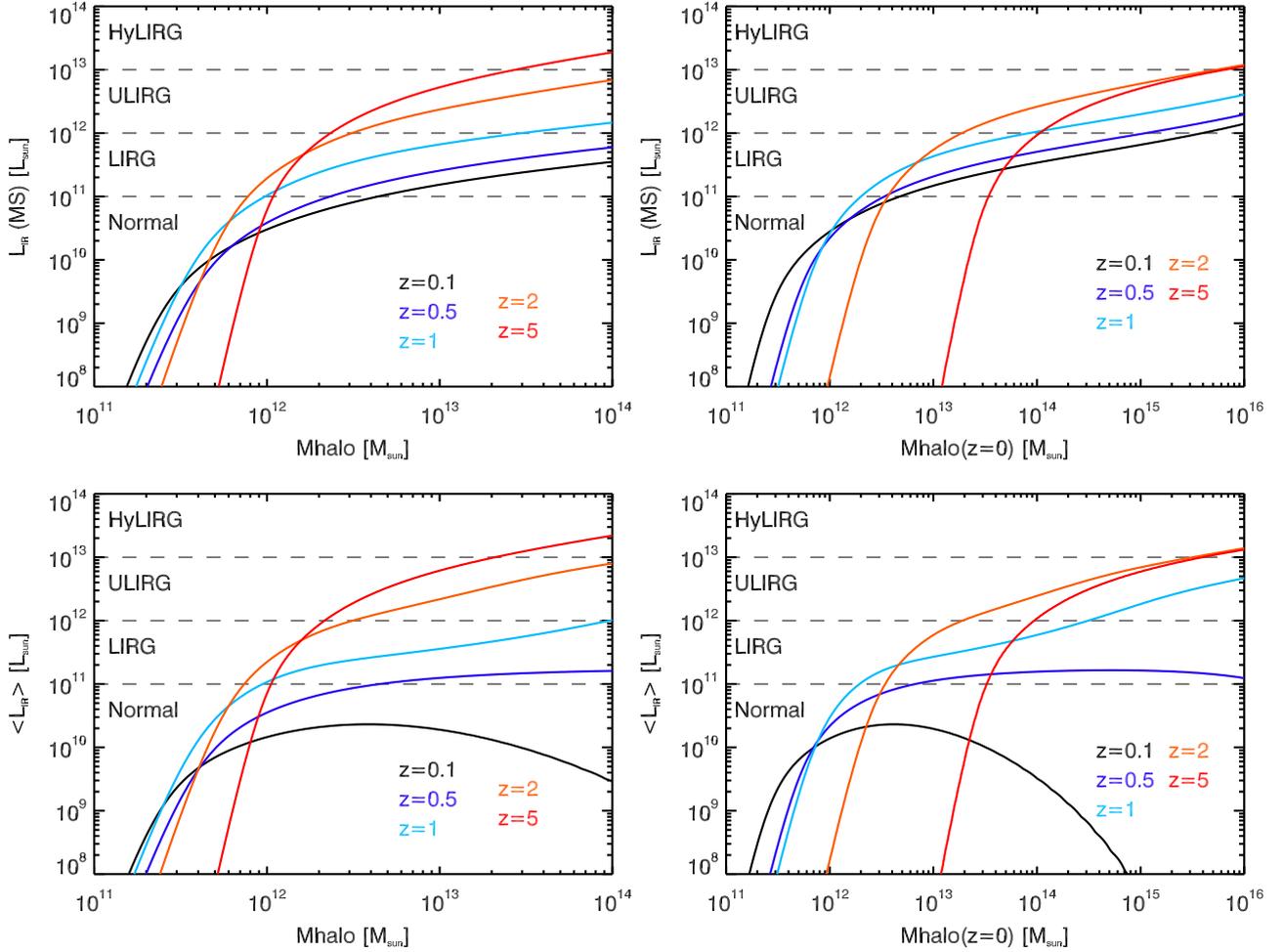}
\caption{\label{fig:LIRMh} Relation between infrared luminosity of galaxies and their halo mass. The \textit{top} panels correspond to the infrared luminosity for a central galaxy exactly on the main sequence. The \textit{bottom} panels shows the mean infrared luminosity of the central galaxies. The \textit{left} panels show results as the instantaneous halo mass and \textit{right panel} as a function of halo mass at z=0. The dashed lines correspond to the limit between normal galaxies (i.e. $<10^{11}$\,L$_{\odot}$), LIRGs ($10^{11}<L_{\rm IR}<10^{12}$\,L$_{\odot}$), ULIRGs ($10^{12}<L_{\rm IR}<10^{13}$\,L$_{\odot}$), and HyLIRGs ($>10^{13}$\,L$_{\odot}$).}
\end{figure*}

\begin{figure*}
\centering
\includegraphics[width=17cm]{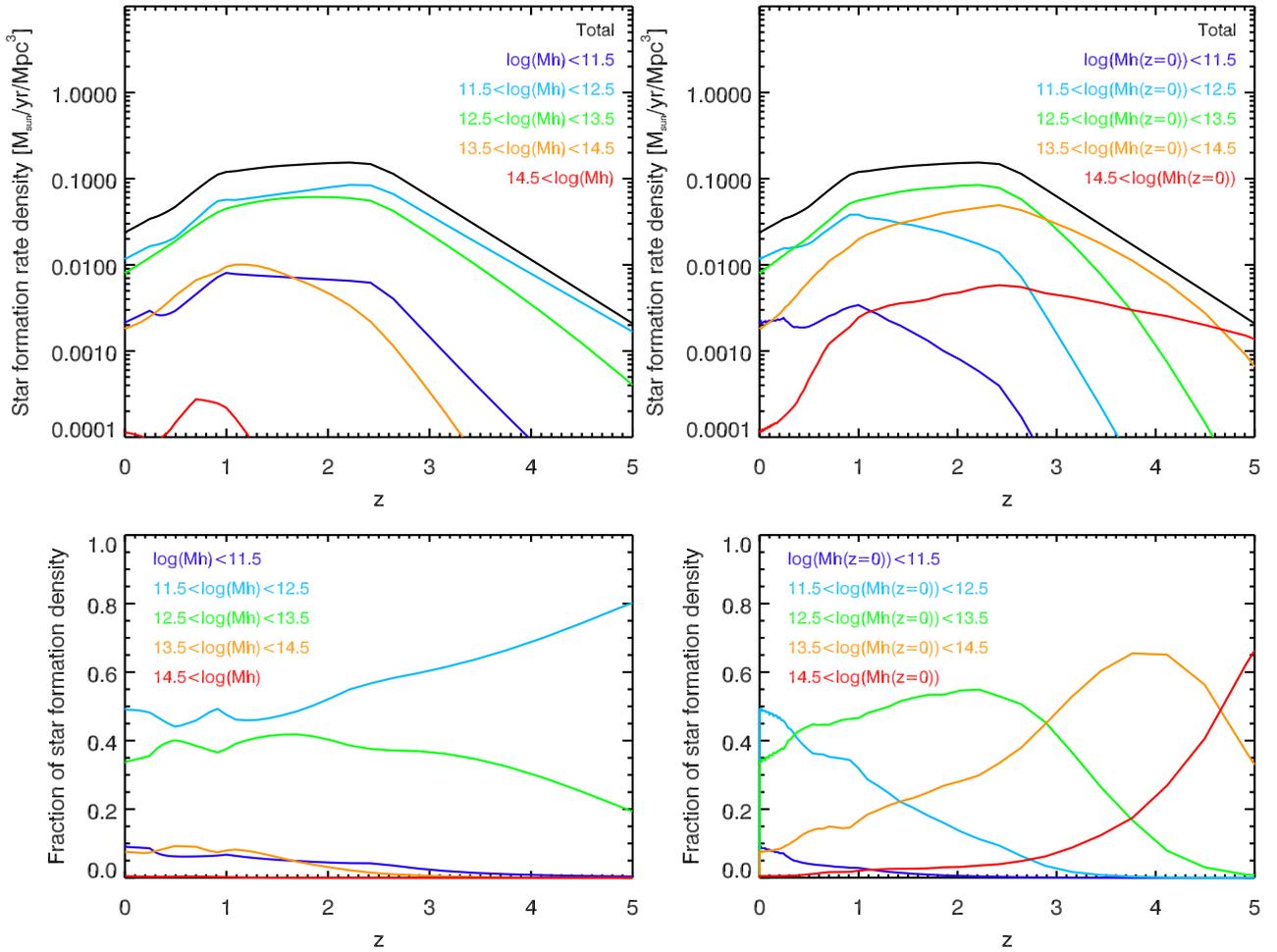}
\caption{\label{fig:madau} Contribution of various halo mass to the star formation history. \textit{Upper panels} shows the contribution of each slice to star formation density and \textit{lower panels} their fractional contribution. We use slice of instantaneous halo mass in the \textit{left panels} and mass at z=0 in the \textit{right panels}.}
\end{figure*}

\begin{figure*}
\centering
\includegraphics[width=16.5cm]{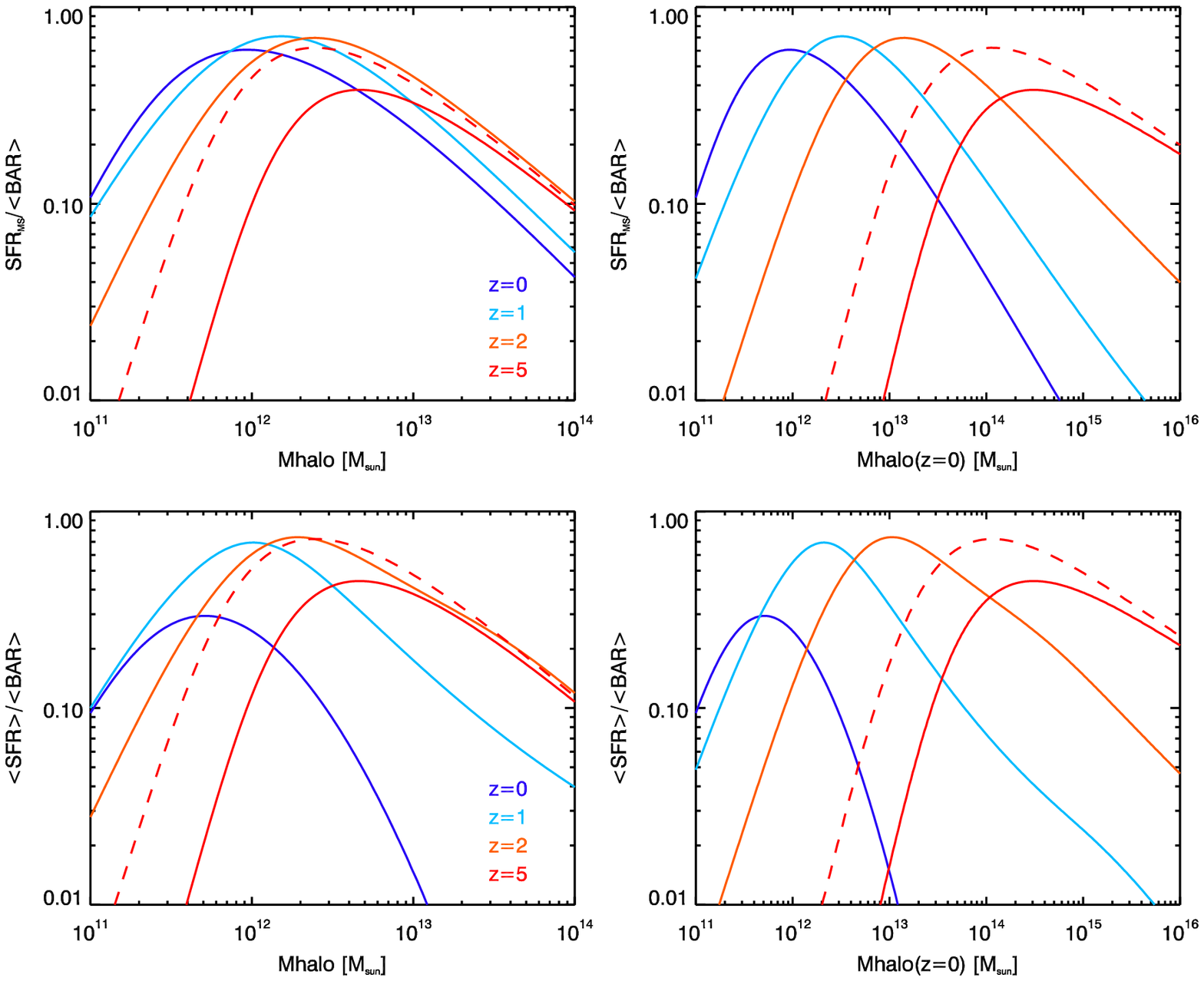}
\caption{\label{fig:sfe} Instantaneous star formation efficiency (defined here as SFR/BAR, where BAR is the baryonic accretion rate) as a function of halo mass at various redshift predicted by our model. The predictions from model C are plotted with a solid line. At $z=5$ model A predicts a different efficiency distribution and is represented by a dashed line. \textit{Upper panels} shows this efficiency only for main-sequence galaxies and \textit{lower panels} for mean efficiency. We use slice of instantaneous halo mass in the \textit{left panels} and mass at z=0 in the \textit{right panels}.}
\end{figure*}

\section{Star formation and dark matter halos}

\label{sect:sfh}

In this section, we discuss the prediction of our model in the context of our current understanding of galaxy formation. We use model C, as discussed in Sect.\,\ref{sect:validity}.\\

\subsection{Evolution of infrared-light-to-mass ratio with redshift}

The infrared-light-to-halo-mass ratio is a key ingredient of our CIB model. Figure\,\ref{fig:LIRMh} shows this relation for various hypotheses. We consider two different definitions of the infrared luminosity inside the halos: the infrared luminosity of a central galaxy lying exactly at the core of the main-sequence (top panels) and the mean infrared luminosity of central galaxies (lower panels) hosted by halos of a given mass. This second quantity takes into account that a fraction of galaxies are passive, while the first quantity only takes the star-forming galaxies into account. We also use both the instantaneous halo mass (left panels) and the halo mass at z=0. The conversion between instantaneous halo mass and halo mass at z=0 is performed assuming an accretion following the fits of \citet{Fakhouri2010} to their own numerical simulations of dark matter:
\begin{equation}
\label{eq:acc}
\begin{split}
\langle \dot{M_h} \rangle = 46.1 \times \textrm{M$_\odot$ / yr$^{-1}$} \left ( \frac{M_h}{10^{12} \textrm{M$_\odot$}}\right )^{1.1} \times (1+1.11 z) \\
\times \sqrt{\Omega_m (1+z)^3 + \Omega_\Lambda}.
\end{split}
\end{equation}
The mass at z=0 {\large (}$M_h(z) + \int_{t(z'=z)}^{t(z'=0)} \langle \dot{M_h} \rangle \, dt$ {\large)} is a convenient quantity to use when discussing the evolution of structures with time.\\

The shape of the infrared luminosity as a function of instantaneous halo mass evolves little with redshift. However, its normalization varies strongly. For the same halo mass, infrared luminosity is much more intense at high redshift. This results will be interpreted further in Sect.\,\ref{sect:sfe}. At low mass, the infrared luminosity decreases quickly. This is mainly caused by the quickly dropping M$_\star$-$\mathcal M_h$ relation at $\mathcal M_h<10^{12}\,$M$_\odot$(see Fig.\,\ref{fig:mstarmh}), while the sSFR is roughly the same at all stellar masses (sSFR$\propto$M$_\star^{-0.2}$). This break is slightly amplified by the fact that a smaller fraction of UV light from young stars is reprocessed into infrared emission at lower masses (Eq.\,\ref{eq:att}). At high mass, we found a sublinear relation, even if we consider only objects on the main sequence. This behavior is also driven by the shape of the M$_\star$-$\mathcal M_h$ relation, which is sublinear at M$_h>$10$^{12}$\,M$_\odot$. This trend is stronger if we consider the mean infrared luminosity and take the quenched galaxies into account. At z$<$1, the infrared luminosity decreases when halo mass becomes higher than $10^{13}$\,M$_\odot$, so our model predicts a very weak star formation in a dense environment at low redshift, in agreement with, say \citet{Feruglio2010}. This trend is caused by the large fraction of quenched galaxies at high stellar mass (see Fig.\,\ref{fig:am}). The contribution of satellites is small except in massive halos ($>10^{13}$\,M$_\odot$), where the central is inefficient in forming stars.\\

The infrared luminosity as a function of halo mass at z=0 exhibits strong downsizing. The progenitors of $10^{14}$\,M$_\odot$ initially host a very intense star formation rate (ULIRG regime, $10^{12}<L_{\rm IR}<10^{13}$\,L$_{\odot}$) at high redshift (z=5), which continues to grow up to z$\sim$2. Then, the infrared luminosity decreases because star formation is less efficient in massive halos (see Sect.\,\ref{sect:sfe}). The less massive halos need more time to ignite intense star formation. Dark matter halos similar to that of MilkyWay ($\sim$10$^{12}$\,M$_\odot$ at z=0) host significant infrared emission only below z=1.\\

We can compare the predictions of our model with estimates based on clustering of various samples of galaxies. Submillimeter galaxies (defined here to have $S_{850}>3$\,mJy) are essentially ULIRGs at z$\sim$2.4 \citep[e.g.][]{Chapman2003} and the B12 model predicts that these objects generally lie on the main sequence. \citet{Blain2004} show that the correlation length of SMGs is compatible with typical host halos of $10^{13}\,M_\odot$, which is consistent with our model prediction of an infrared luminosity of $4\times10^{12}$\,L$_\odot$ for a main-sequence galaxy in a 10$^{13}$\,M$_\odot$ halo at z=2.4 (see Fig.\,\ref{fig:LIRMh}). We can also check the consistency of our results with measurements based on \textit{Spitzer} 24\,$\mu$m observations. \citet{Magliocchetti2008} measured the clustering of two samples of sources with $S_{24}>400$\,$\mu$Jy with a mean redshift of 0.79 and 2.02, respectively. This flux selection corresponds, at the mean redshift of each sample, to an infrared luminosity of $2.6\times10^{11}$\,L$_\odot$ and $2.6\times10^{12}$\,L$_\odot$, respectively for a main-sequence SED. For an object exactly on the main sequence, this corresponds to an instantaneous halo mass of $2.7\times10^{12}$\,M$_\odot$ and $1.2\times10^{13}$\,M$_\odot$, in agreement with the minimal mass found by \citet{Magliocchetti2008} of $0.8_{-0.7}^{+2.3}\times10^{12}$\,M$_\odot$ and  $0.6_{-0.3}^{+0.6}\times10^{13}$\,M$_\odot$, respectively. \citet{Farrah2006} measured the halo mass of  5.8\,$\mu$m IRAC peakers with 24\,$\mu$m detections and found a 1$\sigma$ range for log($M_h$) of 13.7-14.1. This population of massive star-forming galaxies has a mean redshift of 2.017 and a mean infrared luminosity of 8.9$\pm$0.6\,L$_\odot$ \citep{Fiolet2010}, which is associated in our model to log($M_h$)=14.1.\\

\subsection{Contribution of various halo masses to star formation history}

The total star-formation rate of both central and satellite galaxies hosted by a given halo can be derived from our model. In combination with the halo mass function, we can then compute the contribution of each halo mass to the star formation rate. The results of this analysis are presented in Fig.\,\ref{fig:madau}. If we consider instantaneous halo mass (left panels), the bulk of the star formation is hosted by halos between $10^{11.5}$\,M$_\odot$ and $10^{13.5}$\,M$_\odot$ regardless of the redshift, suggesting the existence of a characteristic mass which favors star formation. The fractional contribution of each mass slice is inferred to have evolved slowly with redshift, except at $z>3$. However, there are significant uncertainties in the behavior of the galaxies at high redshift, and these results should be taken with caution.\\

Because of the growth of dark matter halos, the same halo mass at various redshifts corresponds to progenitors of different types of halos. To take this into account, we also considered slices of z=0 halo mass (right panels). The star formation rate density is predicted to have been successively dominated by progenitors of massive clusters ($M_h>10^{14.5}$\,M$_\odot$) at $z>4.7$, small clusters and large group ($10^{13.5}<M_h<10^{14.5}$\,M$_\odot$) at $2.8<z<4.7$, small groups ($10^{12.5}<M_h<10^{13.5}$\,$M_\odot$) at $0.5<z<2.8$, and Milky Way-like halos ($10^{11.5}<M_h<10^{12.5}$\,M$_\odot$) at $z<0.5$. Star formation thus initially occurs in the progenitor of the most massive halos before becoming less efficient and then propagating to less massive halos. This strong downsizing explains why the star formation rate densities at redshifts 1 and 2 are similar: $z>2$ infrared luminosity density is dominated by a low density ($\sim3\times10^{-3}$\,Mpc$^{-4}$) of ULIRGs and $z\sim1$ is dominated by higher density ($\sim3\times10^{-3}$\,Mpc$^{-3}$) of LIRGs \citep[e.g.][]{Bethermin2011}.\\

\subsection{Efficiency of star formation as a function of halo mass and redshift}

\label{sect:sfe}

To conclude, we derive an estimate of the instantaneous star formation efficiency (ISFE) in different halos by computing the ratio between SFR and baryonic accretion rate (BAR). Here, we assume the BAR to be the total matter accretion as given by \citet{Fakhouri2010} multiplied by the universal baryonic faction (BAR = $\langle \dot{M_h} \rangle \times \Omega_b / \Omega_m$ with $\langle \dot{M_h} \rangle$ defined in Eq.\,\ref{eq:acc}). This simplified definition neglects that the gas is not transformed instantaneously into stars and that large gas reservoirs are present in \citep[e.g.,][]{Daddi2010b,Tacconi2012} and around \citep{Cantalupo2012} high-z galaxies. Figure\,\ref{fig:sfe} shows predictions for the variation in this efficiency as a function of halo mass at various redshifts. Star-formation efficiency in main-sequence galaxies as a function of instantaneous halo mass (upper left panel) slowly evolves between z=0 and z=2 with a slight increase in the mass of maximum efficiency with redshift (from $8\times10^{11}$\,M$_\odot$ at z=0 to $3\times10^{12}$\,M$_\odot$ at z=2). At z=5, this maximal efficiency is lower, and the halo mass where it occurs is higher. The increase in the mass of maximum efficiency could be due to a delay in the ignition of star-formation activity, because the BAR in $\sim$10$^{12}$\,M$_\odot$ halos is higher at high redshift. However, large uncertainties exist at high redshift where we have few constraints on star-forming galaxies. For instance, if we use model A (flat sSFR at $z>2.5$) instead of model C (rising sSFR) to perform our calculations this effect is much smaller (dashed line in Fig.\,\ref{fig:sfe}). In our framework, this results is a consequence of the combined evolution of the sSFR, the M$_\star$-M$_h$ relation, and the specific halo growth. We can also try to interpret this evolution physically. Below the mass scale at which peak efficiency is reached, the gravitational potential of the halo is lower and supernova feedback is probably sufficiently strong to remove gas from the galaxy \citep[e.g.][]{Silk2003,Bertone2005}. At higher masses, the slow decrease in the star formation efficiency could be caused by the transition from cold streams below $M_h \sim 10^{12}$\,M$_\odot$ to isotropic cooling above this mass \citep[e.g.][]{Dekel2006,Faucher2011}. At high mass, the cooling time of gas becomes much longer than the free fall time because of the hot atmosphere in the massive halos \citep[e.g.][]{Keres2005,Birnboim2007}. Unlike the ISFE in main sequence galaxies, the mean ISFE (lower left panel), that considers that a fraction of galaxies are quenched, exhibits a strong break at $M_h>10^{12}\,M_\odot$ at $z=0$ and a moderate break at the same mass at $z=1$. This break could be caused by a suppression of isotropic gas cooling by energy injection in the halo atmosphere by active galactic nuclei activity \citep[e.g.][]{Cattaneo2006,Somerville2008,Ostriker2010}.\\

Regardless of the redshift and halo mass, the SFR in main-sequence galaxies is always lower than the BAR onto the dark matter halos (see Fig.\,\ref{fig:sfe}). For a \citet{Chabrier2003} IMF, the maximal efficiency is $\sim$0.7. Appendix\,\ref{app:sfe_sal} discusses the case of a \citet{Salpeter1955} IMF. This is consistent with the standard assumption that main-sequence galaxies host secular star formation. This is not the case for episodic starbursts, which on average forms four times more stars for the same halo mass than main-sequence galaxies and thus transform much more gas into stars than they receive from cosmic accretion. They thus tend to exhaust their gas reservoir rapidly. The main-sequence galaxies do not need to store gas to fuel their star formation. This could seem to contradicts the fact that the specific halo growth (sHG) is larger by a factor of two than the sSFR around $z=2$, as pointed out by \citet{Weinmann2011}. However, the ratio between sSFR and sHG is
\begin{equation}
\frac{\textrm{SFR}/M_\star}{\dot{M}_h/M_h} = \frac{\eta \dot{M}_b / \dot{M}_h}{M_\star / M_h} = \eta \frac{M_b}{M_\star},
\end{equation}
where $\eta$ is the instantaneous star formation efficiency (SFR/$\dot{M}_b$), and $\dot{M}_b$ the baryonic accretion into the halo. The ratio between sSFR and sHG can thus be much higher than unity in halos of $\sim10^{12}$M$_\odot$, because even if $\eta$ is lower than 1, the ratio between baryonic mass and stellar mass is $\sim$4 \citep[e.g.,][]{Leauthaud2012}, because of the low efficiency of conversion of baryons into stars in the past (when the halos had a lower mass and thus a lower ISFE).\\

We can also discuss how the efficiency varies as a function of z=0 halo mass (right panels). We predict a strong evolution of the typical halos where the ISFE is maximal. The progenitors of massive halos form stars very efficiently at high redshift, while Milky-Way-like halos are very inefficient. The opposite trend is expected at low redshift. This picture is consistent with the strong downsizing of the star-forming galaxy population discussed in the previous sections, but also with the work of \citet{Behroozi2012a}. Their analysis was based only on the evolution of the SMF, and SFR were derived assuming a single galaxy population. Our approach is based on the infrared observations that directly probe the SFR in galaxies and which take the diversity of galaxies into account using three distinct populations: secularly star-forming galaxies on the main-sequence, episodic, merger-driven starbursts, and passive elliptical galaxies. Compared to \citet{Behroozi2012b}, our mean ISFE is much lower in local massive halos, which almost exclusively host quenched galaxies, but we agree with their estimate if we take only a main-sequence population (more consistent with their single galaxy population).\\

Normal spiral galaxies at $z=0$, LIRGs at $z=1$, and ULIRGs at $z=2$ dominate the star-formation density at these redshift and are essentially main sequence galaxies \citep{Sargent2012}. All of these are hosted by halos of similar mass, which are characterized by a very efficient conversion of accreted baryons into stars. The huge difference between their star formation rate can thus be explained by the accretion, which is stronger at high redshift. These objects can thus be viewed as different facets of the same universal process of secular star formation.

\section{Conclusion}

\label{sect:conclusion}

We have studied the connection between star formation and dark matter halos focusing on infrared observations. We developed a new modeling approach based on the 2SFM framework, which was already able to successfully reproduce the infrared luminosity function and number counts \citep[][B12]{Sargent2012}. This framework links stellar mass with star formation and infrared properties, when assuming two different modes of star formation in secularly star-forming galaxies and episodic starbursts. We extended this formalism to the connection between stellar mass and halo mass using the technique of abundance matching. Our formalism accounts for the facts that a significant fraction of massive galaxies are passive and do not form stars (especially at low redshift).
\begin{itemize}
\item We developed a method of computing the CIB anisotropies (including power-spectra between different frequencies), the cross-correlation between CIB and CMB lensing, and auto-correlation functions of bright resolved galaxies using the prescription of the 2SFM formalism. To perform this computation, we produced effective SEDs of all galaxies at a given redshift, and these are available online\footnote{\url{http://irfu.cea.fr/Sap/Phocea/Page/index.php?id=537} or CDS via anonymous ftp to cdsarc.u-strasbg.fr (130.79.128.5) or via http://cdsweb.u-strasbg.fr/cgi-bin/qcat?J/A+A/}.
\item We find that a slowly rising sSFR at $z>2.5$ and a quenching of satellite of massive passive galaxies at low redshift (our model C) matches the infrared data reasonably well suggesting that this model is a valid description of the link between infrared galaxies and dark matter halos. The other versions of the model with flat sSFR and no quenching of satellites show some small discrepancies with the power-spectra, but cannot be conclusively ruled out.
\item Our model is able to predict the redshift distribution of CIB anisotropies. We found that the mean redshift where the CIB is emitted varies strongly between 100\,$\mu$m and 850\,$\mu$m but only a little at longer wavelengths. Consequently, the CIB anisotropies in the various bands above 850\,$\mu$m are strongly correlated ($>0.9$).
\item We found a quick rise in the far-infrared-light-to-mass ratio with redshift in $M_h>10^{12}\,M_\odot$ halos and a strong break at lower mass at all redshifts. We found that more than 90\% of the star formation is hosted by halos with masses between $10^{11.5}$ and $10^{13.5}$\,M$_\odot$ at all redshifts. The progenitors of  clusters ($M_h(z=0)>10^{13.5}\,M_\odot$) host the bulk of the star formation at $z>$3. Star formation activity then propagates to groups ($10^{12.5}<M_h(z=0)<10^{13.5}\,M_\odot$) at $0.5<z<3$ and Milky Way-like halos ($10^{11.5}<M_h(z=0)<10^{12.5}\,M_\odot$) at $z<0.5$. We also found that there is a characteristic halo mass ($\sim10^{12}\,M_\odot$) where the star formation efficiency is maximal ($\sim$70\% at all redshift). The large difference of SFR in galaxies dominating the background at low and high redshift would thus be driven by a difference of accretion rate in the halos close to this mass.\\
\end{itemize}

Our simple modeling framework is very efficient in explaining the current observations of the infrared Universe. However, future large submillimeter surveys (e.g. NIKA, CCAT) will resolve the bulk of the CIB into individual sources and will probably improve measurements of the clustering properties of infrared galaxies. We expect that deviation from our model will appear on small scales, where environmental effects could have a strong impact, thereby revealing a more complex and varied infrared Universe.\\

\begin{acknowledgements}
We thank Steve Maddox for providing data and discussion about clustering measurements, Marco Viero and Amir Hajian for providing data, Paolo Serra for useful discussion, and the anonymous referee for very constructive comments. MB, ED, and MS acknowledge the support provided by the grants ERC-StG UPGAL 240039 and ANR-08-JCJC-0008. LW acknowledges support from an ERC StG grant (DEGAS-259586). This research was carried out in part at the Jet Propulsion Laboratory, run by the California Institute of Technology under a contract from NASA.
\end{acknowledgements}

\bibliographystyle{aa}

\bibliography{biblio}

\begin{appendix}

\section{Conversions tables}

\label{sect:conv}

\subsection{Wavenumber, multipole, and angle}

On small scales (\textit{Spitzer}, \textit{Herschel}), where curvature of the sky is negligible, people generally use wavenumber (k) in their measurements of the CIB power spectrum. This is not the case for large-scale measurements (SPT, ACT, \textit{Planck}) for which people used mutlitpole ($\ell$). The conversion between the two is just $\ell = 2 \pi k$. This also corresponds to a characteristic angular scale $\theta=\pi/\ell$. Table\,\ref{tab:angconv} provides conversion for the range of values used in this paper.

\begin{table}
\centering
\caption{\label{tab:angconv} Conversion between wavenumber, multipole, and angle}
\begin{tabular}{ccc}
\hline
\hline
Multipole & Wavenumber & Angle \\
$\ell$ & $k$ & $\theta$ \\
\hline
 & arcmin$^{-1}$ & degree \\
\hline
1 & 0.00005 & 180.00000\\
2 & 0.00009 & 90.00000\\
5 & 0.00023 & 36.00000\\
10 & 0.00046 & 18.00000\\
20 & 0.00093 & 9.00000\\
50 & 0.00231 & 3.60000\\
100 & 0.00463 & 1.80000\\
200 & 0.00926 & 0.90000\\
500 & 0.02315 & 0.36000\\
1000 & 0.04630 & 0.18000\\
2000 & 0.09259 & 0.09000\\
5000 & 0.23148 & 0.03600\\
10000 & 0.46296 & 0.01800\\
20000 & 0.92593 & 0.00900\\
50000 & 2.31481 & 0.00360\\
100000 & 4.62963 & 0.00180\\
\hline
\end{tabular}
\end{table}

\subsection{From wavelengths to frequencies}

Infrared astronomers use wavelengths in $\mu$m, the CMB community frequencies in GHz. For quick reference, we provide the conversion between these two conventions for the passbands discussed in this paper in Table\,\ref{tab:wavefreq}.

\begin{table}
\centering
\caption{\label{tab:wavefreq} Conversion between wavelength and frequency for various passbands used in this paper}
\begin{tabular}{ccc}
\hline
\hline
Wavelength & Frequency \\
\hline
$\mu$m & GHz\\
\hline
24 & 12500 \\
100 & 3000 \\
160 & 1875 \\
250 & 1200 \\
350 & 857\\
500 & 600\\
550 & 545 \\
850 & 353 \\
1100 & 272 \\
1382 & 217\\
2097 & 143\\
3000 & 100 \\
\hline
\end{tabular}
\end{table}

\section{Computation of the cross-Poisson term}

\label{sect:compcp}

The number of sources per steradian $n_{ij}$ with an infrared luminosity in the interval [$L_{\rm IR,i}, L_{\rm IR,i}+\Delta L_{\rm IR,i}$] and a redshift in [$z_j,z_j+\Delta z_j$] is
\begin{equation}
n_{ij} = \frac{dV}{dz} \times \frac{d^2 N}{dL_{\rm IR} dV} \times \Delta L_{\rm IR,i} \Delta z_j.
\end{equation}
We also need to separate the galaxies by mode of star formation (main-sequence or starburst) and $\langle U \rangle$ parameter. The number of main-sequence or starburst galaxies $n_{ijk}^{\rm MS \, or \, SB}$ in the same $L_{\rm IR}$ and z bins and with $\langle U \rangle$ in [$\langle U \rangle_k,\langle U \rangle_k+\Delta \langle U \rangle_k$] is
\begin{equation}
n_{ijk}^{\rm MS \, or \, SB} = \frac{dV}{dz} \times \frac{d^2 N}{dL_{\rm IR} dV} \times \Delta L_{\rm IR,i} \Delta z_j \times p_{\rm MS\, or \, SB}(\langle U \rangle) \times \Delta \langle U \rangle_k.
\end{equation}
Since we assume Poisson statistics, the variance on the number of galaxies in this bin equals $n_{ijk}^{\rm MS \, or \, SB}$. For the covariance between the fluxes at the two frequencies caused by this subpopulation we have
\begin{equation}
\sigma_{S_\nu S_{\nu'},ijk}^{\rm MS\, or \, SB} = S_\nu \times S_{\nu'} \times n_{ijk} = L_{\rm IR,i}^2 s_\nu^{\rm MS\, or \, SB} \times s_{\nu'}^{\rm MS\, or \, SB} \times n_{ijk},
\end{equation}
because the fluxes in the two bands are perfectly correlated for sources with the same SED. Finally, we sum over the entire population (all $L_{\rm IR}$, z, and $\langle U \rangle$) to compute the level of the Poisson term:
\begin{equation}
C_{\ell,\nu \nu'}^{\rm poi} = \sum_{\rm \{MS, SB\}} \sum_{L_{\rm IR, i}} \sum_{z_j} \sum_{\langle U \rangle_k} \sigma_{S_\nu S_{\nu'},ijk}^{\rm MS\, or \, SB},
\end{equation}
which in integral limit becomes
\begin{equation}
\begin{split}
C_{\ell,\nu \nu'}^{\rm poi} =  \int_z \frac{dV}{dz} \sum_{\rm \{MS, SB\}} \int_{\langle U \rangle} p_{\rm MS\, or \, SB} \left(\langle U \rangle|z \right )  \int_{L_{\rm IR}=0}^{L_{\rm IR,cut}^{\rm MS \, or \, SB}(\langle U \rangle,z)} \\
\frac{d^2N_{\rm MS\, or \, SB}}{dL_{\rm IR} dV} L_{\rm IR}^2 s_\nu^{\rm MS\, or \, SB}(\langle U \rangle,z) \times s_{\nu'}^{\rm MS\, or \, SB}(\langle U \rangle,z)\,  dL_{\rm IR} \, d\langle U \rangle \, dz.
\end{split}
\end{equation}

\begin{figure*}
\centering
\includegraphics[width=17cm]{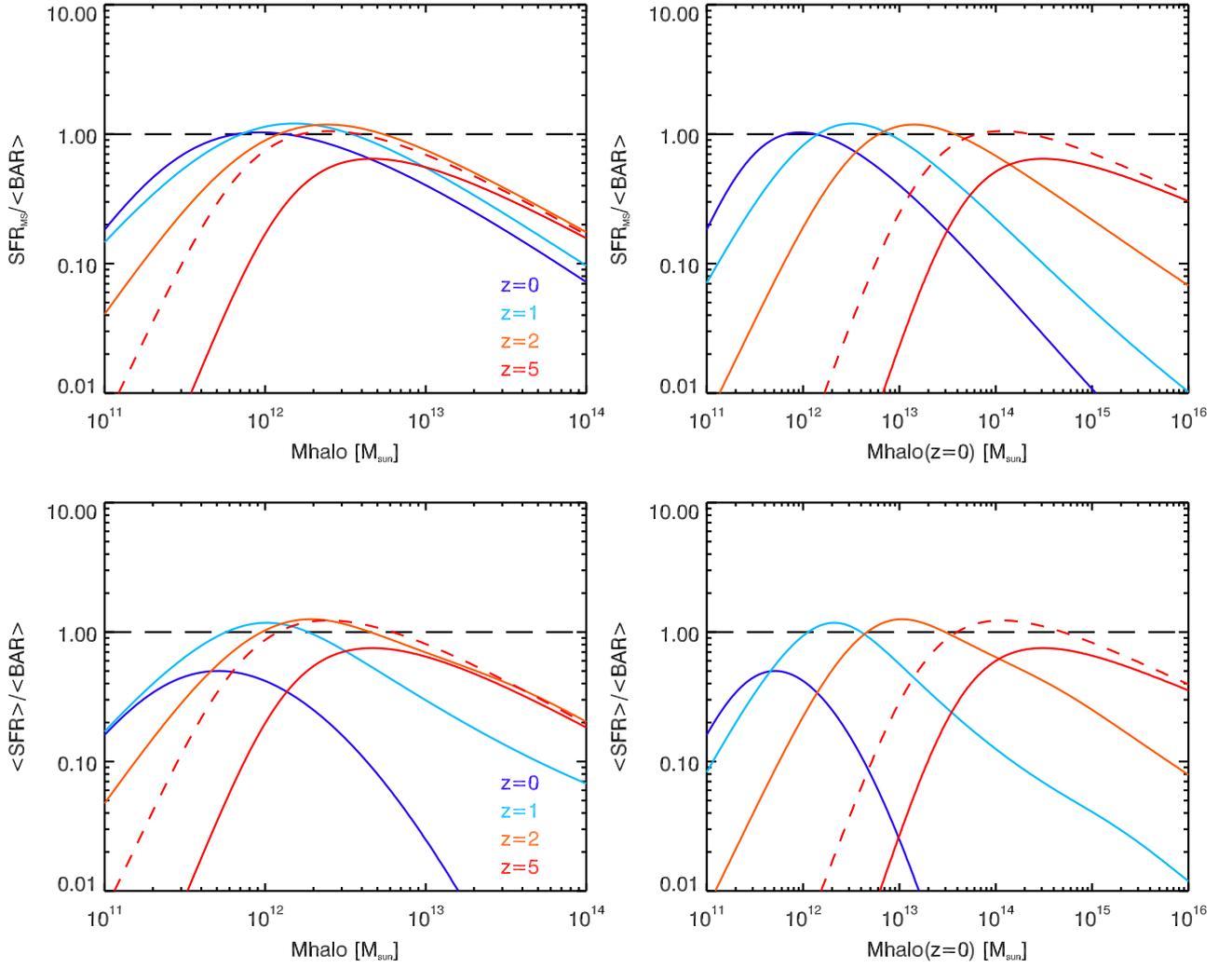}
\caption{\label{fig:sfe_sal} Same figure as \ref{fig:sfe} but assuming a \citet{Salpeter1955} IMF. The black dashed line corresponds to an SFE of 1.} 
\end{figure*}

\section{Star formation efficiency in the case of a Salpeter IMF}

\label{app:sfe_sal}

In Sect.\,\ref{sect:sfe}, we computed the SFE assuming a \citet{Chabrier2003} IMF. The results are slightly different if we assume a \citet{Salpeter1955} IMF (see Fig.\,\ref{fig:sfe_sal}). In this case, the galaxies close to the mass of maximum ISFE form more stars than they accrete baryons, and large gas reservoirs are required to allow the secular star formation in these objects. These reservoirs could have been replenished during the phase of low star-formation efficiency, when the halo was less massive.\\

\end{appendix}

\end{document}